\documentclass{article}

\usepackage{PRIMEarxiv}

\usepackage[utf8]{inputenc} % allow utf-8 input
\usepackage[T1]{fontenc}    % use 8-bit T1 fonts
\usepackage{hyperref}       % hyperlinks
\usepackage{url}            % simple URL typesetting
\usepackage{booktabs}       % professional-quality tables
\usepackage{amsfonts}       % blackboard math symbols
\usepackage{nicefrac}       % compact symbols for 1/2, etc.
\usepackage{microtype}      % microtypography
\usepackage{lipsum}
\usepackage{fancyhdr}       % header
\usepackage{graphicx}       % graphics
\graphicspath{{media/}}     % organize your images and other figures under media/ folder
\usepackage{url,hyperref,lineno,microtype}
\usepackage[onehalfspacing]{setspace}
\usepackage{amsmath,amsfonts,amssymb}
\usepackage{float}
\usepackage{lmodern}
\usepackage{natbib}
\usepackage[resetlabels]{multibib}
\usepackage{url,hyperref,lineno,microtype,subcaption}
\usepackage{caption}

\usepackage{booktabs}
\newtheorem{proposition}{Proposition}
\newtheorem{remark}{Remark}

\usepackage{etoolbox}

\newif\iflegacyfrontierssubfig

\renewenvironment{subfigure}[1][]{%
    \legacyfrontierssubfigtrue
    \ifstrempty{#1}
        {\begin{figure}}
        {\begin{figure}[#1]}%
}{%
    \end{figure}%
    \legacyfrontierssubfigfalse
}

\AtBeginEnvironment{minipage}{%
    \iflegacyfrontierssubfig
        \setcaptionsubtype
    \fi
}

\newcites{app}{Appendix References}

\title{Voxel-wise Bayesian Estimation for Multi-Population Positronium Lifetime Imaging
}

\author{%
\parbox{0.95\textwidth}{%
\centering
\large\bfseries
Berkin Uluutku$^{1}$,
Narendra Rathod$^{2}$,
Giulianno Gasparato$^{1}$,
Katrina Stephenson$^{1}$,
Axel Rominger$^{2}$,
Kuangyu Shi$^{2}$,
Hsin-Hsiung Huang$^{1,*}$
\\[0.8em]
\normalfont\small
$^{1}$School of Data, Mathematical, and Statistical Sciences,
University of Central Florida,
Orlando, Florida, United States of America
\\[0.35em]
$^{2}$Department of Biomedical Research,
University of Bern, and University Clinic for Nuclear Medicine,
Inselspital Bern, Switzerland
\\[0.6em]
$^{*}$Corresponding author:
\texttt{hsin-hsiung.huang@ucf.edu}
}%
}

\begin{document}
\maketitle

\begin{abstract}
Positronium lifetime imaging (PLI) provides information on the local annihilation environment beyond conventional activity imaging, but existing approaches often estimate lifetime parameters over predefined regions and may not account for multiple lifetime populations within the same spatial location. In this work, we present a three-dimensional population-specific Bayesian framework for fast voxel-wise PLI. A partial system matrix describes the spatial probability of each detected event, while measured lifetimes provide soft assignments to slow, fast, and noise populations. These population-specific event responsibilities are then used to estimate a decay-rate posterior independently for each supported voxel, preserving local lifetime variation together with statistical uncertainty.
The framework was evaluated using simulated and experimental data. In simulation, the population-specific formulation recovered spatially varying slow-population decay rates and a common fast-population rate, while a single-population model produced systematic bias. Slow-population two-standard-deviation coverage ranged from 93.8\% to 98.1\% across the simulated regions. Experimental validation using $^{124}$I triple-coincidence data acquired on a Siemens Biograph Vision Quadra scanner produced separate slow- and fast-population lifetime maps for aluminum, nickel, copper, and quartz. The long-lived quartz component was consistent with ortho-positronium, while the fast population showed material-dependent differences among the metal samples. Fast-population coverage was lower (69.1\%), indicating that the fast-population uncertainty remains underestimated. The method is computationally efficient, requiring less than ten seconds per population on a single CPU core requiring only seconds to minutes per population for voxel-wise lifetime estimation across both datasets. The proposed framework therefore provides fast, population-specific, voxel-wise PLI with Bayesian uncertainty quantification, offering a practical statistical imaging approach for spatially heterogeneous lifetime estimation. This combination makes spatially resolved statistical inference feasible without sacrificing computational practicality for volumetric PLI applications.
\end{abstract}

% keywords can be removed
\keywords{Bayesian inference \and multi-population modeling \and positronium lifetime imaging \and positron emission tomography \and voxel-wise estimation}

\section{Introduction}
Positron emission tomography (PET) provides three-dimensional images of radiotracer distribution and is widely used to characterize physiological and metabolic processes in vivo. Conventional PET primarily derives contrast from the spatial distribution of detected annihilation events and therefore provides limited information about the physical processes occurring between positron emission and annihilation. During this interval, a positron may form positronium, a short-lived bound state of a positron and an electron. Positronium properties depend on interactions with the surrounding medium, and its lifetime has been associated with characteristics of the local microenvironment, including free volume, molecular environment, and oxygen concentration \citep{shibuya_oxygen_2020,moskal_positronium_2022}. This sensitivity has motivated positronium lifetime imaging (PLI) as an extension of PET in which the time between positron emission and annihilation provides an additional source of image contrast \citep{moskal_feasibility_2019,moskal_positronium_2021}. Experimental developments have progressed from feasibility and phantom studies to measurements with clinical PET systems and, more recently, in vivo and voxel-wise imaging \citep{steinberger_positronium_2024, moskal_positronium_2024, mercolli_vivo_2026}. These studies demonstrate that positronium lifetime can be used as a spatial imaging quantity complementary to conventional PET activity. As PLI moves toward broader three-dimensional and clinical applications, an important objective is to recover local lifetime behavior throughout the imaging volume so that spatial variations in the annihilation environment can be identified and evaluated directly.

Voxel-wise lifetime estimation is important because regional or sample-level analyses summarize all events within a selected volume using a small set of lifetime parameters. Such estimates are valuable for characterizing an overall material or tissue response, but spatial variation within the region is lost through aggregation. In contrast, voxel-wise estimation treats lifetime as an image parameter and can preserve local differences, gradients, and boundaries within the measured volume. This is particularly relevant for biological imaging, where the annihilation environment may vary within a lesion or surrounding tissue and a single regional lifetime may not represent that heterogeneity. Voxel-wise lifetime maps can also be evaluated directly alongside other spatially resolved information, including PET activity and anatomical imaging. Statistical reconstruction methods have therefore increasingly considered lifetime estimation directly in image space \citep{qi_positronium_2022,huang_statistical_2025}, and recent work has demonstrated that voxel-wise PLI can be performed in vivo using clinical PET/CT data \citep{mercolli_vivo_2026}. These developments establish voxel-wise lifetime estimation as a practical direction for PLI and motivate methods that can retain this spatial information while providing a more complete statistical description of the measured lifetime signal.

Several statistical reconstruction methods have been developed to estimate positronium lifetime directly in image space. Early work established maximum-likelihood and penalized maximum-likelihood formulations for voxel-wise lifetime estimation from TOF-PET measurements, demonstrating that spatial lifetime information can be recovered beyond the localization imposed directly by the scanner timing resolution \citep{qi_positronium_2022}. Subsequent approaches have explored different strategies for improving the computational efficiency, spatial resolution, and robustness of lifetime reconstruction. SPLIT reformulated the reconstruction problem using statistical time-thresholding, providing an alternative to direct lifetime-parameter estimation \citep{huang_split_2024}. More recently, SIMPLE was developed as a fast high-resolution reconstruction strategy for positron lifetime tomography, further demonstrating that lifetime information can be recovered at spatial resolutions finer than those obtained through direct event localization alone \citep{huang_fast_2025}. Together, these developments have established a growing family of reconstruction methods for recovering spatially resolved lifetime information from TOF-PET data, with increasing emphasis on computational efficiency and practical image formation.

Our earlier work introduced a probabilistic reconstruction framework that represented the uncertain spatial origin of each detected event through voxel-wise source probabilities derived from detector geometry and TOF information, enabling voxel-wise lifetime estimation in two dimensions \citep{huang_statistical_2025}. This framework was also extended to measurements containing multiple positron populations. Similarly, a two-component reconstruction approach explicitly modeled slow- and fast-decay populations and showed that population separation can reduce the bias produced by a single-component lifetime model \citep{chen_enhanced_2024}. These multi-population approaches demonstrated the feasibility of resolving distinct lifetime components spatially, but were developed in two dimensions and relied on iterative numerical estimation of the voxel-wise lifetime parameters. More recently, we extended voxel-wise lifetime estimation to fully three-dimensional data using a partial system matrix and a conjugate Bayesian formulation \citep{uluutku_conjugate_2026}. This approach enabled computationally efficient voxel-wise posterior estimation together with local uncertainty information, but represented the lifetime distribution in each voxel using a single effective exponential component. Thus, efficient three-dimensional voxel-wise estimation and multi-population lifetime modeling have so far been addressed separately rather than within a unified statistical framework.

The use of a single effective lifetime becomes limiting when several annihilation processes contribute to the measured lifetime distribution. Experimental PLI measurements can contain distinct short- and long-lived components together with background or mismatched events, and recent analyses on clinical PET systems have shown that multi-component decay behavior must be considered when interpreting measured lifetime spectra \citep{steinberger_positronium_2024}. When these contributions are represented by a single decay rate, the resulting estimate reflects a mixture of the underlying populations and may obscure differences associated with the individual lifetime components. Separating the populations is therefore particularly important in voxel-wise PLI, where the objective is not only to identify that different lifetime scales are present, but also to determine how each population varies spatially throughout the imaging volume. Extending population separation to voxel-wise three-dimensional estimation introduces two related sources of uncertainty. The spatial origin of a detected event is not known exactly and must be distributed probabilistically among possible source voxels, while the measured lifetime likewise does not uniquely identify the population from which the event originated. A population-specific voxel-wise framework must therefore account for both spatial and lifetime information, allowing each event to contribute fractionally to multiple candidate voxels and lifetime populations. This provides a path toward separate spatial maps of the underlying lifetime components rather than a single effective lifetime map, while retaining events whose spatial or population assignment is inherently uncertain.

A fully Bayesian formulation is attractive for voxel-wise PLI because it provides more than a point estimate of the lifetime parameter. By representing the decay rate probabilistically, Bayesian inference can provide a posterior distribution for each voxel and population, allowing local uncertainty to be quantified directly and carried alongside the estimated lifetime maps. This is particularly valuable in PLI, where the number of events contributing to individual voxels can vary substantially and where both spatial and population assignments are uncertain. In principle, a full Bayesian model could jointly account for these latent assignments and lifetime parameters. However, posterior inference for such a model generally requires iterative sampling or numerical approximation over a large number of voxel-wise parameters. The computational burden becomes especially important in three-dimensional imaging, where thousands to hundreds of thousands of voxels may each require population-specific inference. A practical Bayesian PLI framework must therefore retain the statistical advantages of posterior inference while avoiding the computational cost of performing a separate high-dimensional numerical inference problem at every voxel.

In this work, we extend three-dimensional voxel-wise Bayesian PLI to multiple lifetime populations. The proposed framework combines probabilistic spatial information with soft lifetime-based population assignments, allowing each detected event to contribute fractionally to both candidate source voxels and candidate lifetime populations. Population-specific event weights are then used in a conjugate Bayesian formulation to obtain a separate decay-rate posterior for each supported voxel and population, together with voxel-wise uncertainty estimates, without requiring numerical posterior sampling. In this way, the framework retains the spatial resolution of voxel-wise PLI while separating distinct lifetime components and preserving statistical information about the local estimates. The resulting computation remains efficient because, once the spatial and population responsibilities are established, the posterior update for each voxel and population is available in closed form and can be evaluated independently.

The proposed method is evaluated using both simulated and experimental data. Simulation studies are used to assess population separation, voxel-wise decay-rate recovery, and uncertainty coverage under controlled conditions. Experimental validation is performed using $^{124}$I triple-coincidence data acquired on a Siemens Biograph Vision Quadra PET scanner, with aluminum, nickel, copper, and quartz samples providing distinct measured lifetime behavior. Together, these studies evaluate whether population-specific Bayesian inference can provide fast, spatially resolved lifetime estimates with local uncertainty in realistic three-dimensional PLI. The goal is to move beyond both regional multi-component fitting and single-population voxel-wise estimation toward a practical framework for population-specific statistical imaging.

\section{Methods}
This section presents the proposed three-dimensional multi-population PLI reconstruction framework. We first describe the list-mode event data and the preprocessing used to obtain the detector-time channels and measured lifetimes. We then construct the partial system matrix, estimate the detected activity distribution, and calculate the spatial probability that each event originated from each candidate voxel. These spatial probabilities are divided among the ortho-positronium, fast, and noise populations using the measured event lifetimes. The population-specific weights are then used to obtain voxel-wise decay-rate posteriors through conjugate Bayesian inference. Finally, we describe the simulated and experimental datasets used to evaluate the reconstruction.

\subsection{List-mode PLI Event Representation}
The reconstruction begins with a list of detected triple-coincidence events. Each event contains two annihilation photons and one prompt gamma photon. For event $k$, we write

\begin{equation}
e_k
=
\left(
d_{1,k},
d_{2,k},
t_{1,k},
t_{2,k},
d_{p,k},
t_{p,k}
\right),
\qquad
k=1,\ldots,N_{\mathrm{ev}},
\label{eq:list_mode_event}
\end{equation}

where $d_{1,k}$ and $d_{2,k}$ denote the detectors that record the two annihilation photons, and \(d_{p,k}\) denotes the detector that records the prompt gamma. The corresponding detection times are $t_{1,k}$, $t_{2,k}$, and $t_{p,k}$.

The two annihilation photons are ordered according to their detection times. We define \(t_{1,k}\) as the first annihilation-photon detection and $t_{2,k}$ as the second, such that

\begin{equation}
t_{1,k}\leq t_{2,k}.
\end{equation}

The ordered annihilation-detector pair is then

\begin{equation}
\eta_k
=
\left(d_{1,k},d_{2,k}\right).
\label{eq:detector_pair}
\end{equation}

The difference between the annihilation-photon detection times provides the time-of-flight information for the event,

\begin{equation}
\Delta t_k
=
t_{2,k}-t_{1,k}.
\label{eq:tof_difference}
\end{equation}

The continuous time difference is assigned to a discrete TOF bin \(\xi_k\) with width \(\Delta t_{\mathrm{bin}}\). The detector pair and TOF bin together define the detector-time channel

\begin{equation}
c_k
=
\left(\eta_k,\xi_k\right).
\label{eq:detector_time_channel}
\end{equation}

Events recorded by the same ordered detector pair and assigned to the same TOF bin belong to the same channel. Let $y_c$ denote the number of retained events in channel $c$. The observed channel count is

\begin{equation}
y_c
=
\sum_{k=1}^{N_{\mathrm{ev}}}
\mathbb{I}(c_k=c),
\label{eq:channel_count}
\end{equation}

where $\mathbb{I}(\cdot) $is the indicator function. The set of observed channels is

\begin{equation}
\mathcal{C}_{+}
=
\left\{
c:y_c>0
\right\}.
\label{eq:observed_channels}
\end{equation}

Only these observed channels are included in the partial system matrix. The total number of retained events therefore satisfies

\begin{equation}
N_{\mathrm{ev}}
=
\sum_{c\in\mathcal{C}_{+}}y_c.
\label{eq:retained_event_count}
\end{equation}

The event lifetime is calculated from the prompt and annihilation detection times. The annihilation time is estimated as the average detection time of the two annihilation photons, orrected for the flight time of the annihilation photons,

%\begin{equation}
%t_{a,k}
%=
%\frac{t_{1,k}+t_{2,k}}{2}.
%\label{eq:annihilation_time}
%\end{equation}

\begin{equation}
t_{a,k}
=
\frac{t_{1,k}+t_{2,k}}{2}
-\;\frac{D_{\eta_k}}{2c}.
\label{eq:annihilation_time}
\end{equation}

Here $D_{\eta_k}$ is the distance between the two annihilation detectors and $c$ is the speed of light. Because the two annihilation-photon path lengths always sum to $D_{\eta_k}$, this correction is exact and depends only on the detector pair, not on the emission position along the line of response.

The measured prompt-to-annihilation time difference is then obtained after subtracting the flight time of the prompt gamma,

%\begin{equation}
%\tau_k
%=
%t_{a,k}-t_{p,k}.
%\label{eq:measured_lifetime}
%\end{equation}

\begin{equation}
\tau_k
=
t_{a,k}
-\left(
t_{p,k}
-\;\frac{\lVert\widehat{\boldsymbol{r}}_k-\boldsymbol{r}_{d_{p,k}}\rVert}{c}
\right)
,
\label{eq:measured_lifetime}
\end{equation}

where $\boldsymbol{r}_{d_{p,k}}$ is the position of the prompt-gamma detector and $\widehat{\boldsymbol{r}}_k$ is the estimated emission point of event $k$. Unlike the annihilation-photon term, this correction depends on the emission position and therefore requires a point estimate; we use the TOF-estimated position along the line of response.

Events are retained only if they satisfy the detector, energy, coincidence, and lifetime selection criteria used for the corresponding dataset. The lifetime is restricted to the selected interval

\begin{equation}
\tau_{\min}
\leq
\tau_k
\leq
\tau_{\max}.
\label{eq:lifetime_interval}
\end{equation}

The acquisition-specific selection criteria and retained event counts are provided in Sections~2.6 and~2.7 for the simulated and experimental datasets, respectively.

After preprocessing, each retained event is represented by its detector-time channel and measured lifetime,

\begin{equation}
\mathcal{D}
=
\left\{
(c_k,\tau_k)
\right\}_{k=1}^{N_{\mathrm{ev}}}.
\label{eq:processed_event_data}
\end{equation}

The detector-time channel $c_k$ provides the spatial information used to assign the event to candidate voxels. The measured lifetime $\tau_k$ is used to divide this spatial assignment among the lifetime populations and estimate their voxel-wise decay rates.

\subsection{Spatial Event-Voxel Assignment}
The detector pair and TOF measurement restrict the possible origin of an event but do not determine a single source voxel. Because the scanner has finite timing resolution, each observed detector-time channel is generally compatible with several voxels. We represent this uncertainty using the partial system matrix introduced in our previous work. A conventional full system matrix contains a row for every indexed detector-pair and TOF-bin combination supported by the scanner model. It therefore includes channels that are not present in the acquired data and channels that provide no useful support within the selected reconstruction volume. In three dimensions, the number of these possible channels can be very large. Constructing and storing the complete matrix is therefore computationally expensive, even though only a fraction of its rows may contribute to a given acquisition. Furthermore, although the available channels may form a complete set of emission outcomes in an idealized two-dimensional closed-ring geometry, a finite three-dimensional scanner does not enclose the reconstruction volume, and photons can leave through its axial ends without producing a detected channel.

The partial system matrix is constructed only for the observed channels,

\begin{equation}
\mathcal{C}_{+}
=
\left\{
c:y_c>0
\right\}.
\end{equation}

This restriction reduces the matrix to the detector-time channels that contribute to the measured data. More importantly, each row is formulated as a conditional probability over the possible source voxels of an event that has already been detected.

Let \(J\) denote the source voxel, \(C=c\) the observed detector-time channel, and \(D=1\) the detection of a valid event. The reconstruction volume is

\begin{equation}
\mathcal{V}
=
\left\{
1,\ldots,N_{\mathrm{vox}}
\right\}.
\end{equation}

As described in our previous work, the unnormalized elements of the partial system matrix are calculated by combining Siddon ray tracing with the Gaussian TOF response \cite{chen_enhanced_2024,huang_statistical_2025,uluutku_conjugate_2026,siddon_fast_1985}. Siddon's algorithm determines the segment of the line of response contained within each voxel. If \(s_{c,j}^{-}\) and \(s_{c,j}^{+}\) are the corresponding entry and exit positions, the unnormalized weight is

\begin{equation}
\widetilde{H}_{c,j}
=
\int_{s_{c,j}^{-}}^{s_{c,j}^{+}}
g_c(s)\,ds,
\label{eq:unnormalized_partial_system_matrix}
\end{equation}

where \(g_c(s)\) is the Gaussian TOF response for channel \(c\). The weights are then normalized over the reconstruction volume,

\begin{equation}
H_{c,j}
=
\frac{
\widetilde{H}_{c,j}
}{
\displaystyle
\sum_{\ell=1}^{N_{\mathrm{vox}}}
\widetilde{H}_{c,\ell}
}.
\label{eq:partial_system_matrix_normalization}
\end{equation}

Under an equal prior probability for the candidate voxels, the normalized element can be interpreted as

\begin{equation}
H_{c,j}
=
P\left(
J=j
\mid
C=c,\,
D=1,\,
J\in\mathcal{V}
\right).
\label{eq:partial_system_matrix_probability}
\end{equation}

The probabilities in Equation~\ref{eq:partial_system_matrix_normalization} are calculated from the physical measurement model of the scanner. The annihilation-detector pair defines the line of response, while the TOF measurement provides information about the source position along that line. The calculation also incorporates the photon paths through the reconstruction volume and the relevant timing and detector-response physics. Voxels that are incompatible with the measured line of response receive zero probability. Among the remaining voxels, those that agree more closely with the measured TOF position receive larger values.

The observed data determine which rows are retained, but they do not determine the values within those rows. These values are fixed by the physical scanner model. The partial system matrix therefore introduces the scanner geometry, photon propagation, and timing response directly into the spatial assignment of each event.

Because the event is conditioned on originating within the reconstruction volume, each row of the partial system matrix is normalized over its candidate source voxels,

\begin{equation}
\sum_{j=1}^{N_{\mathrm{vox}}}
H_{c,j}
=
1,
\qquad
c\in\mathcal{C}_{+}.
\label{eq:partial_system_matrix_row_sum}
\end{equation}

Thus, \(H_{c,j}\) is not the probability that an emission from voxel \(j\) will be detected. Detection and the measured channel are already included as conditions. Instead, \(H_{c,j}\) gives the probability that an event observed in channel \(c\) originated from voxel \(j\), before information from the detected activity distribution is included.

The observed channel counts are next used to reconstruct the spatial distribution of the retained detected events. Let \(\widetilde{f}_j\) denote the detected event count associated with voxel \(j\). This quantity represents the number of retained detected events attributed to the voxel within the reconstruction model, rather than the total number of physical decays in that voxel.

Starting from a positive estimate \(\widetilde{f}_j^{(n)}\), the predicted event count in channel \(c\) is

\begin{equation}
\widehat{y}_c^{(n)}
=
\sum_{j=1}^{N_{\mathrm{vox}}}
H_{c,j}\widetilde{f}_j^{(n)}.
\label{eq:predicted_channel_count}
\end{equation}

The detected event distribution is updated by redistributing each observed channel count across its candidate voxels,

\begin{equation}
\widetilde{f}_j^{(n+1)}
=
\widetilde{f}_j^{(n)}
\sum_{c\in\mathcal{C}_{+}}
H_{c,j}
\frac{
y_c
}{
\widehat{y}_c^{(n)}
}.
\label{eq:detected_event_update}
\end{equation}

The update is repeated for a fixed number of iterations, allowing the detected event distribution to stabilize. The final estimate is combined with the partial system matrix to obtain the spatial responsibility for each event. For event \(k\), the corresponding row is \(H_{c_k,j}\), and the activity-informed spatial responsibility is

\begin{equation}
w_{k,j}
=
\frac{
H_{c_k,j}\widetilde{f}_j
}{
\displaystyle
\sum_{\ell=1}^{N_{\mathrm{vox}}}
H_{c_k,\ell}\widetilde{f}_{\ell}
}.
\label{eq:spatial_event_responsibility}
\end{equation}

These responsibilities satisfy

\begin{equation}
\sum_{j=1}^{N_{\mathrm{vox}}}
w_{k,j}
=
1.
\label{eq:spatial_responsibility_sum}
\end{equation}

The value \(w_{k,j}\) represents the spatial probability assigned to voxel \(j\) after accounting for both the physical compatibility encoded by the partial system matrix and the reconstructed distribution of detected events. In the next subsection, this spatial probability is divided among the ortho-positronium, fast, and noise populations using the measured lifetime \(\tau_k\).

\subsection{Population-Specific Event-Voxel Assignment}
The spatial responsibility \(w_{k,j}\) describes the probability assigned to voxel \(j\) for event \(k\), but it does not identify the lifetime population that produced the event. We assume that each active voxel may contain contributions from both slow and fast lifetime populations. Although their decay rates may vary spatially, rates belonging to the same population are assumed to be more similar across voxels than rates belonging to different populations. This separation allows the global lifetime distribution to provide an initial population assignment before the voxel-wise rates are reconstructed.

Let

\begin{equation}
Z_k
\in
\left\{
s,f,n
\right\}
\end{equation}

denote the population assigned to event \(k\), where \(s\), \(f\), and \(n\) represent the slow, fast, and noise populations, respectively. These labels describe the observed lifetime populations without assigning each component to a single physical annihilation process. In particular, the fast population may contain contributions from direct annihilation and short-lived positronium states.

The signal and noise fractions are determined from the measured global lifetime distribution, with the noise assumed to be uniform over the retained lifetime interval. Within the signal population, the slow and fast fractions are fixed at \(0.30\) and \(0.70\), respectively. If \(\pi_n\) denotes the estimated noise fraction, the mixture proportions are

%
% something
%

\begin{equation}
\pi_s
=
0.30(1-\pi_n),
\qquad
\pi_f
=
0.70(1-\pi_n),
\qquad
\pi_s+\pi_f+\pi_n=1.
\label{eq:population_mixture_proportions}
\end{equation}

Because these signal fractions are fixed and applied uniformly to all materials, the slow population is populated even for materials in which no genuine long-lived component physically exists. For the metal samples this is expected to produce an effective slow component that does not correspond to a bulk-metal annihilation state, rather combination of noise and some other physical effects. The resulting effective component is discussed in Section~\ref{sec:experimental_results} The measured global lifetime distribution is modeled using two exponential signal populations and a uniform noise population. The slow and fast densities are

\begin{equation}
f_s(\tau_k\mid\lambda_s)
=
\lambda_s e^{-\lambda_s\tau_k},
\qquad
f_f(\tau_k\mid\lambda_f)
=
\lambda_f e^{-\lambda_f\tau_k},
\label{eq:global_signal_densities}
\end{equation}

where \(\lambda_s\) and \(\lambda_f\) are the global slow and fast decay rates. The noise density is assumed to be uniform over the retained lifetime interval,

\begin{equation}
f_n(\tau_k)
=
\frac{1}{\tau_{\max}-\tau_{\min}},
\qquad
\tau_{\min}
\leq
\tau_k
\leq
\tau_{\max}.
\label{eq:uniform_noise_density}
\end{equation}

The resulting global mixture density is

\begin{equation}
f(\tau_k)
=
\pi_s f_s(\tau_k\mid\lambda_s)
+
\pi_f f_f(\tau_k\mid\lambda_f)
+
\pi_n f_n(\tau_k).
\label{eq:global_lifetime_mixture}
\end{equation}

The mixture proportions and uniform noise density are held fixed. The global slow and fast rates are estimated from all retained lifetimes. Starting from initial values \(\lambda_s^{(0)}\) and \(\lambda_f^{(0)}\), the responsibility denominator at iteration \(t\) is

\begin{equation}
d_k^{(t)}
=
\pi_s\lambda_s^{(t)}
e^{-\lambda_s^{(t)}\tau_k}
+
\pi_f\lambda_f^{(t)}
e^{-\lambda_f^{(t)}\tau_k}
+
\pi_n f_n(\tau_k).
\label{eq:global_responsibility_denominator}
\end{equation}

The population probabilities for event \(k\) are then

\begin{align}
r_{k,s}^{(t)}
&=
\frac{
\pi_s\lambda_s^{(t)}
e^{-\lambda_s^{(t)}\tau_k}
}{
d_k^{(t)}
},
\\
r_{k,f}^{(t)}
&=
\frac{
\pi_f\lambda_f^{(t)}
e^{-\lambda_f^{(t)}\tau_k}
}{
d_k^{(t)}
},
\\
r_{k,n}^{(t)}
&=
\frac{
\pi_n f_n(\tau_k)
}{
d_k^{(t)}
}.
\label{eq:global_population_responsibilities}
\end{align}

These probabilities satisfy

\begin{equation}
r_{k,s}^{(t)}
+
r_{k,f}^{(t)}
+
r_{k,n}^{(t)}
=
1.
\label{eq:population_responsibility_sum}
\end{equation}

The two signal rates are updated using the events assigned to their corresponding populations,

\begin{equation}
\lambda_s^{(t+1)}
=
\frac{
\displaystyle\sum_{k=1}^{N_{\mathrm{ev}}}
r_{k,s}^{(t)}
}{
\displaystyle\sum_{k=1}^{N_{\mathrm{ev}}}
r_{k,s}^{(t)}\tau_k
},
\qquad
\lambda_f^{(t+1)}
=
\frac{
\displaystyle\sum_{k=1}^{N_{\mathrm{ev}}}
r_{k,f}^{(t)}
}{
\displaystyle\sum_{k=1}^{N_{\mathrm{ev}}}
r_{k,f}^{(t)}\tau_k
}.
\label{eq:global_rate_updates}
\end{equation}

The updates are repeated until the global rate estimates become stable. The final global rates are used to calculate the event-level probabilities

\begin{equation}
r_{k,z}
=
P(Z_k=z\mid\tau_k),
\qquad
z\in\{s,f,n\}.
\label{eq:final_population_probabilities}
\end{equation}

These global rates are used only to separate the measured events into lifetime populations. They are not the final voxel-wise decay-rate estimates.

We treat the spatial and lifetime measurements as two separate sources of assignment information. The spatial responsibility \(w_{k,j}\) determines where the event originated, while \(r_{k,z}\) determines which lifetime population produced it. Their product defines the population-specific event-to-voxel responsibility,

\begin{equation}
w_{k,j}^{(z)}
=
w_{k,j}r_{k,z},
\qquad
z\in\{s,f,n\}.
\label{eq:population_specific_spatial_weight}
\end{equation}

The slow, fast, and noise contributions are therefore

\begin{equation}
w_{k,j}^{(s)}
=
w_{k,j}r_{k,s},
\qquad
w_{k,j}^{(f)}
=
w_{k,j}r_{k,f},
\qquad
w_{k,j}^{(n)}
=
w_{k,j}r_{k,n}.
\label{eq:population_specific_weights}
\end{equation}

Because the population probabilities sum to one, the population-specific contributions preserve the original spatial responsibility,

\begin{equation}
w_{k,j}^{(s)}
+
w_{k,j}^{(f)}
+
w_{k,j}^{(n)}
=
w_{k,j}.
\label{eq:population_weight_partition}
\end{equation}

Thus, the spatial probability assigned to each event is divided among the slow, fast, and noise populations rather than replaced or independently normalized. The slow and fast weights are used in the voxel-wise conjugate reconstruction, while the noise weights are retained separately and excluded from both decay-rate updates.

\subsection{Population-Specific Conjugate Reconstruction}

The preceding subsections assign two probabilities to each retained event. Equation~\ref{eq:spatial_event_responsibility} gives the spatial responsibility \(w_{k,j}\), which describes how much of event \(k\) is assigned to voxel \(j\). Equation~\ref{eq:final_population_probabilities} gives the population probability \(r_{k,z}\), which describes how much of the same event is assigned to population \(z\). Their product,

\begin{equation}
w_{k,j}^{(z)}
=
w_{k,j}r_{k,z},
\end{equation}

represents the fraction of event \(k\) assigned jointly to voxel \(j\) and population \(z\). These joint responsibilities connect the spatial and population assignments to the voxel-wise lifetime reconstruction.

In a fully hierarchical Bayesian model, the source voxel \(J_k\) and population \(Z_k\) would both be treated as latent variables. Their joint assignment can be represented by

\begin{equation}
A_{k,j,z}
=
\mathbb{I}
\left(
J_k=j,\,
Z_k=z
\right).
\label{eq:joint_assignment_indicator}
\end{equation}

A fully assigned event has \(A_{k,j,z}=1\) for one voxel-population pair and zero for all others. Joint sampling of these assignments would propagate the spatial and population uncertainty directly into the voxel-wise decay rates. However, this would require sampling the voxel and population assignment of every event together with all voxel-wise parameters.

To retain a closed-form reconstruction, we divide this hierarchy into conditional stages. The spatial and population assignments are first estimated using the methods described in the preceding subsections. The latent joint indicator is then replaced by its assignment probability,

\begin{equation}
E\left[
A_{k,j,z}
\mid
\mathcal{D}
\right]
\approx
w_{k,j}^{(z)}.
\label{eq:expected_joint_assignment}
\end{equation}

Each event therefore contributes fractionally to several voxel-population pairs rather than being assigned to one pair through sampling or a hard decision.

For a known voxel and population, each measured lifetime represents one complete event. The event contributes one count and its full measured lifetime,

\begin{equation}
n_{k,j,z}
=
1,
\qquad
S_{k,j,z}
=
\tau_k.
\end{equation}

Its physical lifetime model is exponential,

\begin{equation}
\tau_k
\mid
\lambda_{j,z}
\sim
\operatorname{Exponential}
\left(
\lambda_{j,z}
\right).
\label{eq:physical_exponential_lifetime}
\end{equation}

In the present reconstruction, however, the source voxel and population are not known with certainty. Let

\begin{equation}
a_{k,j,z}
=
w_{k,j}^{(z)}
\end{equation}

denote the fraction of event \(k\) assigned to voxel \(j\) and population \(z\). This partial event contributes the fractional event count \(a_{k,j,z}\) and the corresponding fractional accumulated lifetime,

\begin{equation}
n_{k,j,z}
=
a_{k,j,z},
\qquad
x_{k,j,z}
=
a_{k,j,z}\tau_k.
\label{eq:partial_event_contribution}
\end{equation}

A complete exponential event is equivalent to a Gamma contribution with an event count of one. The Gamma form also permits a positive fractional event count. We therefore represent the partial event using

\begin{equation}
f\left(
x_{k,j,z}
\mid
\lambda_{j,z},
a_{k,j,z}
\right)
=
\frac{
\lambda_{j,z}^{\,a_{k,j,z}}
}{
\Gamma\left(a_{k,j,z}\right)
}
x_{k,j,z}^{\,a_{k,j,z}-1}
\exp\left(
-\lambda_{j,z}x_{k,j,z}
\right),
\label{eq:fractional_gamma_likelihood}
\end{equation}

for \(a_{k,j,z}>0\). Here, the Gamma shape \(a_{k,j,z}\) has a direct interpretation as the effective fraction of one event assigned to the voxel-population pair. When \(a_{k,j,z}=1\), the event is fully assigned and Equation~\ref{eq:fractional_gamma_likelihood} reduces to the exponential likelihood. When \(a_{k,j,z}=0\), the event makes no contribution. Thus, the physical lifetime model remains exponential. The Gamma form is introduced because uncertain event assignments produce partial event counts rather than complete counts of one.

Since \(a_{k,j,z}\) and \(x_{k,j,z}\) are treated as known at this stage, the terms involving \(\Gamma(a_{k,j,z})\) and \(x_{k,j,z}^{a_{k,j,z}-1}\) are constant with respect to \(\lambda_{j,z}\). The contribution of event \(k\) is therefore

\begin{equation}
f\left(
x_{k,j,z}
\mid
\lambda_{j,z},
a_{k,j,z}
\right)
\propto
\lambda_{j,z}^{\,a_{k,j,z}}
\exp\left(
-\lambda_{j,z}a_{k,j,z}\tau_k
\right).
\label{eq:fractional_gamma_kernel}
\end{equation}

Multiplying over the retained events gives

\begin{equation}
L\left(
\lambda_{j,z}
\right)
\propto
\lambda_{j,z}^{\,n_{j,z}}
\exp\left(
-\lambda_{j,z}S_{j,z}
\right),
\label{eq:population_voxel_likelihood}
\end{equation}

where

\begin{equation}
n_{j,z}
=
\sum_{k=1}^{N_{\mathrm{ev}}}
w_{k,j}^{(z)}
\label{eq:population_effective_count}
\end{equation}

is the effective event count and

\begin{equation}
S_{j,z}
=
\sum_{k=1}^{N_{\mathrm{ev}}}
w_{k,j}^{(z)}\tau_k
\label{eq:population_effective_lifetime}
\end{equation}

is the effective accumulated lifetime. The same assignment probability therefore scales both the event count and its lifetime contribution.

For each voxel \(j\) and signal population \(z\in\{s,f\}\), we use a diffuse Gamma prior with shape \(\alpha_{0,z}\) and rate \(\beta_{0,z}\),

\begin{equation}
\lambda_{j,z}
\sim
\operatorname{Gamma}
\left(
\alpha_{0,z},
\beta_{0,z}
\right).
\label{eq:population_rate_prior}
\end{equation}

The prior parameters are selected to contribute little information relative to the measured events. Combining this prior with Equation~\ref{eq:population_voxel_likelihood} gives the closed-form posterior

\begin{equation}
\lambda_{j,z}
\mid
\mathcal{D},
\boldsymbol{w}^{(z)}
\sim
\operatorname{Gamma}
\left(
\alpha_{0,z}+n_{j,z},
\beta_{0,z}+S_{j,z}
\right).
\label{eq:population_rate_posterior}
\end{equation}

The reconstructed decay rate is taken as the posterior mean,

\begin{equation}
\widehat{\lambda}_{j,z}
=
\frac{
\alpha_{0,z}+n_{j,z}
}{
\beta_{0,z}+S_{j,z}
}.
\label{eq:population_posterior_mean}
\end{equation}

The corresponding lifetime estimate is

\begin{equation}
\widehat{\tau}_{j,z}
=
\frac{
1
}{
\widehat{\lambda}_{j,z}
}.
\label{eq:population_lifetime_estimate}
\end{equation}

Because the posterior mean is available in closed form, no posterior sampling or voxel-wise numerical optimization is required. Once the population-specific assignments are calculated, each slow and fast rate is reconstructed from two weighted sums.

The posterior in Equation~\ref{eq:population_rate_posterior} is conditional on the calculated spatial and population probabilities. Once the population-specific weights \(w_{k,j}^{(z)}\) are obtained, they are treated as known and fixed in the conjugate reconstruction. Consequently, \(n_{j,z}\) and \(S_{j,z}\) are also treated as fixed, and the conditional Gamma variance is

\begin{equation}
\operatorname{Var}_{\mathrm{Gamma}}
\left(
\lambda_{j,z}
\mid
\mathcal{D},
\boldsymbol{w}^{(z)}
\right)
=
\frac{
\alpha_{0,z}+n_{j,z}
}{
\left(
\beta_{0,z}+S_{j,z}
\right)^2
}.
\label{eq:conditional_gamma_variance}
\end{equation}

This conditional variance does not include the uncertainty represented by the assignment probabilities. The value \(w_{k,j}^{(z)}\) is the expected value of the unknown joint assignment indicator \(A_{k,j,z}\), not the realized assignment itself. For a fixed voxel-population pair,

\begin{equation}
A_{k,j,z}
\sim
\operatorname{Bernoulli}
\left(
w_{k,j}^{(z)}
\right),
\end{equation}

and therefore

\begin{equation}
\operatorname{Var}
\left(
A_{k,j,z}
\right)
=
w_{k,j}^{(z)}
\left(
1-w_{k,j}^{(z)}
\right).
\label{eq:joint_assignment_variance}
\end{equation}

The first-order correction propagates the variance of this unknown assignment through the effective event count and accumulated lifetime. It does not assign a variance to the calculated probability \(w_{k,j}^{(z)}\) itself. Instead, it restores the uncertainty in the latent assignment that is lost when the probability is treated as a fixed partial event.

Assuming independence between retained events, the assignment-induced variance of the effective event count is

\begin{equation}
\operatorname{Var}
\left(
n_{j,z}
\right)
=
\sum_{k=1}^{N_{\mathrm{ev}}}
w_{k,j}^{(z)}
\left(
1-w_{k,j}^{(z)}
\right).
\label{eq:effective_count_variance}
\end{equation}

The corresponding variance of the effective accumulated lifetime is

\begin{equation}
\operatorname{Var}
\left(
S_{j,z}
\right)
=
\sum_{k=1}^{N_{\mathrm{ev}}}
w_{k,j}^{(z)}
\left(
1-w_{k,j}^{(z)}
\right)
\tau_k^2,
\label{eq:effective_lifetime_variance}
\end{equation}

and their covariance is

\begin{equation}
\operatorname{Cov}
\left(
n_{j,z},
S_{j,z}
\right)
=
\sum_{k=1}^{N_{\mathrm{ev}}}
w_{k,j}^{(z)}
\left(
1-w_{k,j}^{(z)}
\right)
\tau_k.
\label{eq:effective_count_lifetime_covariance}
\end{equation}

We propagate these terms through the posterior mean in Equation~\ref{eq:population_posterior_mean}. A first-order approximation gives

\begin{align}
\operatorname{Var}_{\mathrm{assign}}
\left(
\widehat{\lambda}_{j,z}
\right)
&=
\frac{
\operatorname{Var}(n_{j,z})
}{
\left(
\beta_{0,z}+S_{j,z}
\right)^2
}
+
\frac{
\left(
\alpha_{0,z}+n_{j,z}
\right)^2
}{
\left(
\beta_{0,z}+S_{j,z}
\right)^4
}
\operatorname{Var}(S_{j,z})
\nonumber
\\
&\quad
-
2
\frac{
\alpha_{0,z}+n_{j,z}
}{
\left(
\beta_{0,z}+S_{j,z}
\right)^3
}
\operatorname{Cov}
\left(
n_{j,z},
S_{j,z}
\right).
\label{eq:assignment_variance_correction}
\end{align}

The final variance estimate combines the conditional Gamma variance with the assignment-induced variance,

\begin{equation}
\operatorname{Var}
\left(
\lambda_{j,z}
\right)
\approx
\operatorname{Var}_{\mathrm{Gamma}}
\left(
\lambda_{j,z}
\right)
+
\operatorname{Var}_{\mathrm{assign}}
\left(
\widehat{\lambda}_{j,z}
\right).
\label{eq:corrected_population_rate_variance}
\end{equation}

This correction does not change the reconstructed posterior mean. It adds the uncertainty associated with the unknown voxel and population assignments that is omitted when their probabilities are treated as fixed. The resulting framework retains the physical exponential lifetime model, represents uncertain assignments as partial Gamma events, and provides closed-form population-specific three-dimensional rate estimates with a first-order correction for assignment uncertainty.

\subsection{Three-Dimensional Simulation Dataset}

The simulated phantom occupies a \(26\times26\times6\) voxel grid with voxel dimensions of \(0.5~\mathrm{cm}\times0.5~\mathrm{cm}\times3~\mathrm{cm}\), giving a total volume of \(13~\mathrm{cm}\times13~\mathrm{cm}\times18~\mathrm{cm}\). It contains five cylindrical disk-shaped inclusions positioned at the vertices of a regular pentagon. The disk centers lie on a circle with a radius of \(5~\mathrm{cm}\), and each disk has a radius of \(1.8~\mathrm{cm}\) and an axial thickness of approximately \(9~\mathrm{cm}\). The disks have equal activity, and no activity is present outside them.

Each simulated event was assigned to either the slow or fast population. The slow and fast fractions were \(0.30\) and \(0.70\), respectively. The five disks were assigned slow-population decay rates of

\begin{equation}
\lambda_s
\in
\left\{
0.3,\,
0.4,\,
0.5,\,
0.6,\,
0.7
\right\}
~\mathrm{ns}^{-1}.
\end{equation}

The fast-population decay rate was spatially constant within all five disks,

\begin{equation}
\lambda_f
=
2.0~\mathrm{ns}^{-1}.
\end{equation}

No uniform noise population was included in the simulation. Therefore, the population assignment used \(\pi_n=0\), \(\pi_s=0.30\), and \(\pi_f=0.70\). The fast-population ground-truth map is not shown because its rate is constant throughout the active regions.

\begin{figure}[H]
\centering
\includegraphics[width=5in]{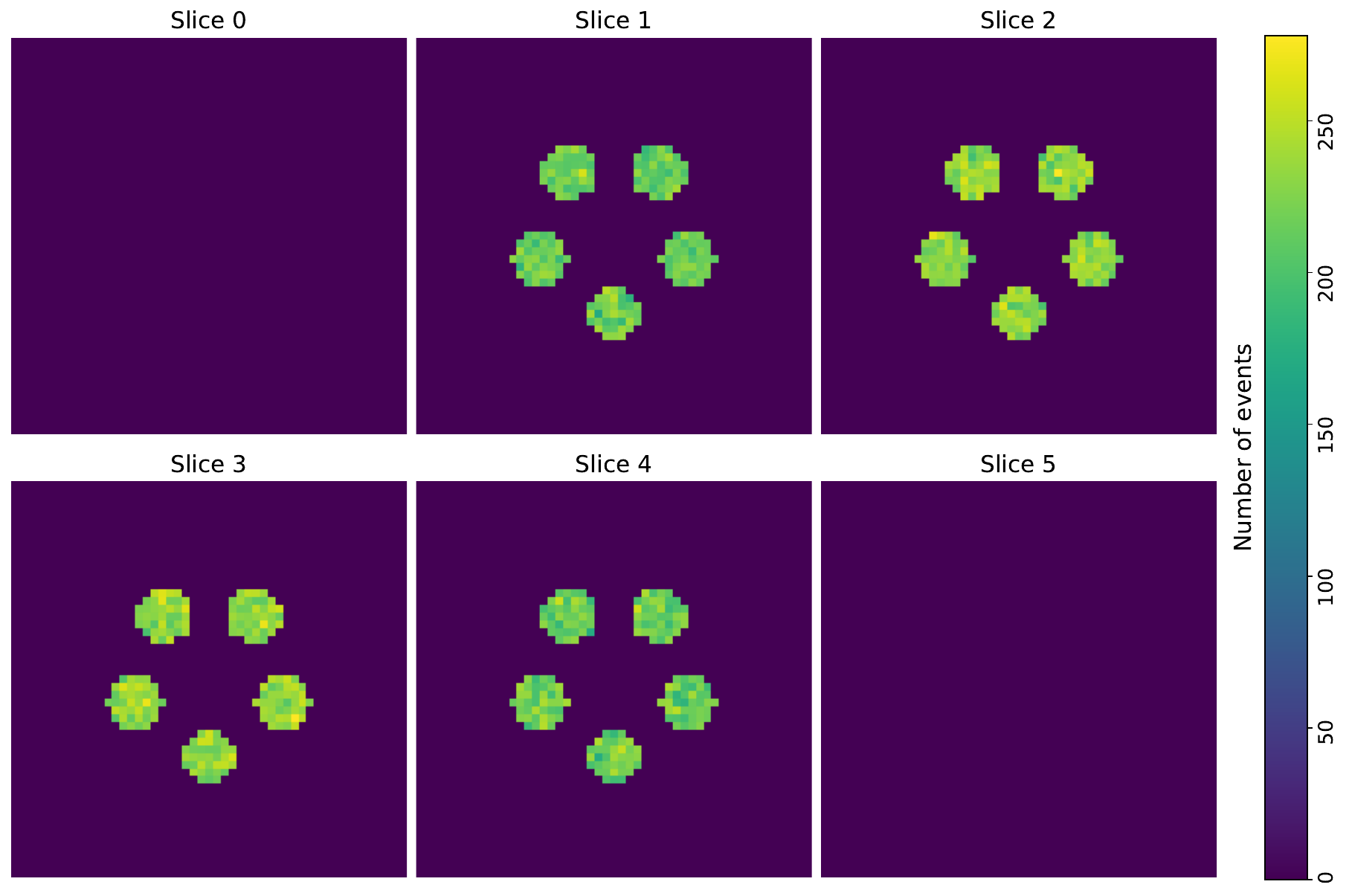}
\caption{
Ground-truth detected activity map of the simulated three-dimensional phantom. The color scale represents the number of retained triple-coincidence events per voxel.
}
\label{fig:phantom_activity}
\end{figure}

\begin{figure}[H]
\centering
\includegraphics[width=5in]{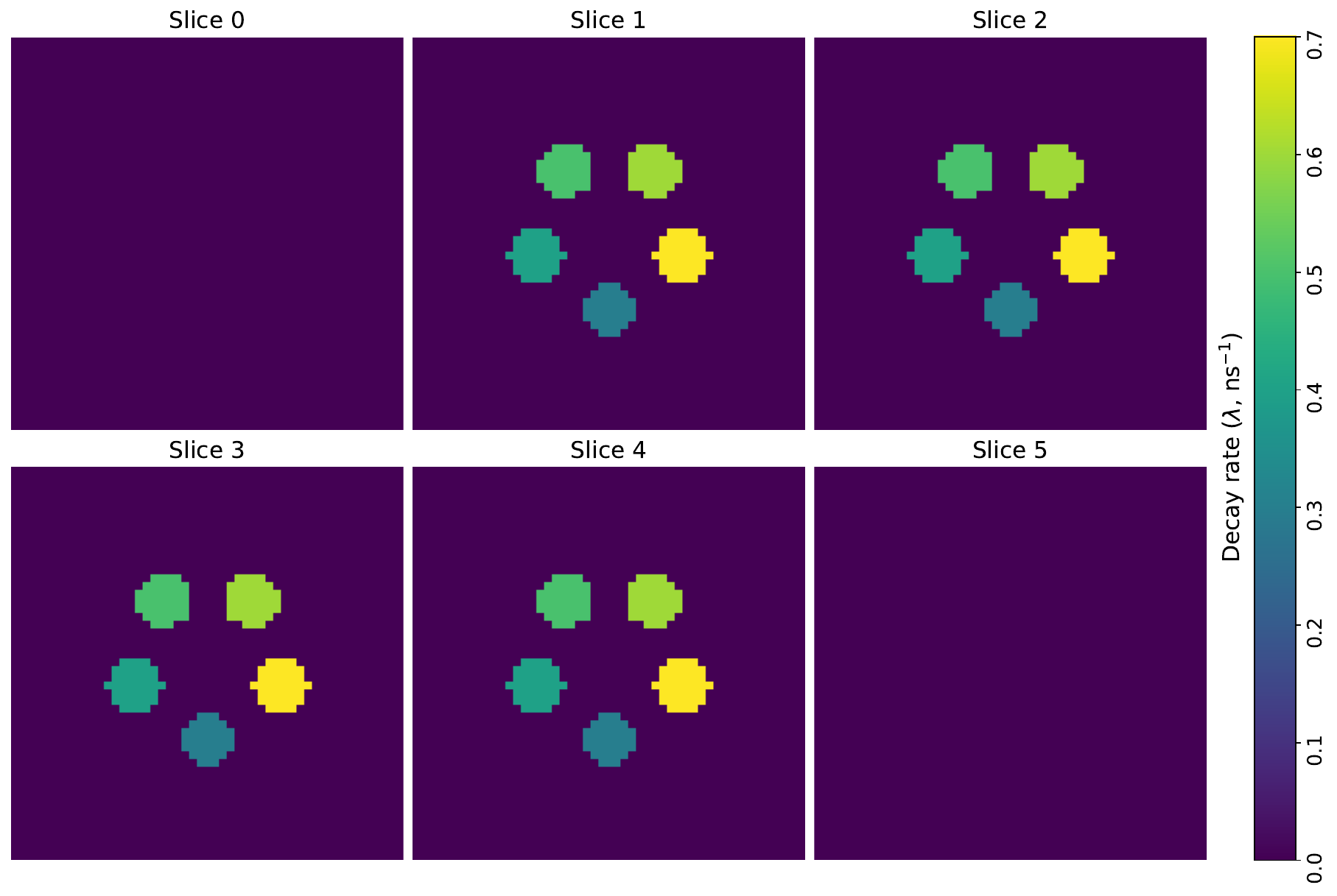}
\caption{
Ground-truth slow-population decay-rate map. The five disk-shaped inclusions have decay rates between \(0.3\) and \(0.7~\mathrm{ns}^{-1}\).
}
\label{fig:phantom_lambda}
\end{figure}

A total of \(1\times10^6\) decays was distributed equally among the five disks. Approximately \(160\times10^3\) triple-coincidence events were detected and retained for reconstruction, a retained triple-coincidence fraction of approximately 16\%. Because the probability of detecting all three photons varies across the scanner, the ground-truth detected activity map represents the retained event population rather than the originally generated decays. The detected activity and slow-population decay-rate maps are shown in Figures~\ref{fig:phantom_activity} and~\ref{fig:phantom_lambda}, respectively.

The simulated scanner is a cylindrical detector array with a diameter of \(60~\mathrm{cm}\) and an axial length of \(50~\mathrm{cm}\). It contains 12 rings with 288 detectors per ring, giving 3456 detector elements. The coincidence timing resolution is \(200~\mathrm{ps}\) FWHM, corresponding to a spatial TOF uncertainty of approximately \(3~\mathrm{cm}\) along the line of response. Simulated event lifetimes were not explicitly blurred by the detector timing response, although annihilation-photon timing was discretized through TOF binning for spatial reconstruction.

\subsection{Experimental Dataset}
Experimental list-mode data acquired using a Siemens Biograph Vision Quadra long-axial-field-of-view PET scanner were used to evaluate the reconstruction on measured events. The acquisition was previously reported in literature ~\cite{lapkiewicz_determining_2024,steinberger_positronium_2024}, where the quartz-glass sample was analyzed using regional lifetime distributions. The present study uses the complete three-dimensional acquisition and includes the metal samples that were not assessed in the original analysis.

The experiment contained paired disks of aluminum, nickel, copper, and quartz glass. A \(^{124}\mathrm{I}\) sodium iodide solution was placed between each pair of disks, and the samples were arranged across the scanner field of view. Triple-coincidence events were formed within a \(20~\mathrm{ns}\) coincidence window. The annihilation-photon energy window was \(460\text{--}545~\mathrm{keV}\), while the \(602.7~\mathrm{keV}\) prompt gamma was selected using an energy window of \(568\text{--}639~\mathrm{keV}\). An additional \(4.2~\mathrm{ns}\) coincidence window was applied between the two annihilation photons. Correlated background events originating from \(^{176}\mathrm{Lu}\) in the detector crystals were reduced by requiring the prompt interaction to be separated from each annihilation interaction by more than 30 crystals, corresponding to approximately \(100~\mathrm{mm}\). After preprocessing, approximately \(2\times10^5\) triple-coincidence events were retained.

The reconstruction volume was discretized into \(125\times125\times12\) voxels. The voxel dimensions were \(0.33~\mathrm{cm}\times0.33~\mathrm{cm}\times1.0~\mathrm{cm}\). The partial system matrix was constructed from the detector geometry and TOF measurements of the retained events. A timing resolution of \(228~\mathrm{ps}\) FWHM was used for the Gaussian TOF response. The resulting detector-time channels and flight-path-corrected lifetime measurements were used as inputs to the population-specific reconstruction.

\section{Results and Discussion}

The proposed framework was evaluated using both simulated and experimental datasets. The results focus on the accuracy of population-specific voxel-wise lifetime reconstruction, the effect of separating multiple lifetime populations, and the uncertainty associated with the resulting estimates.

\subsection{Simulation Results}

\begin{figure}[h!]
\centering
\includegraphics[width=6in]{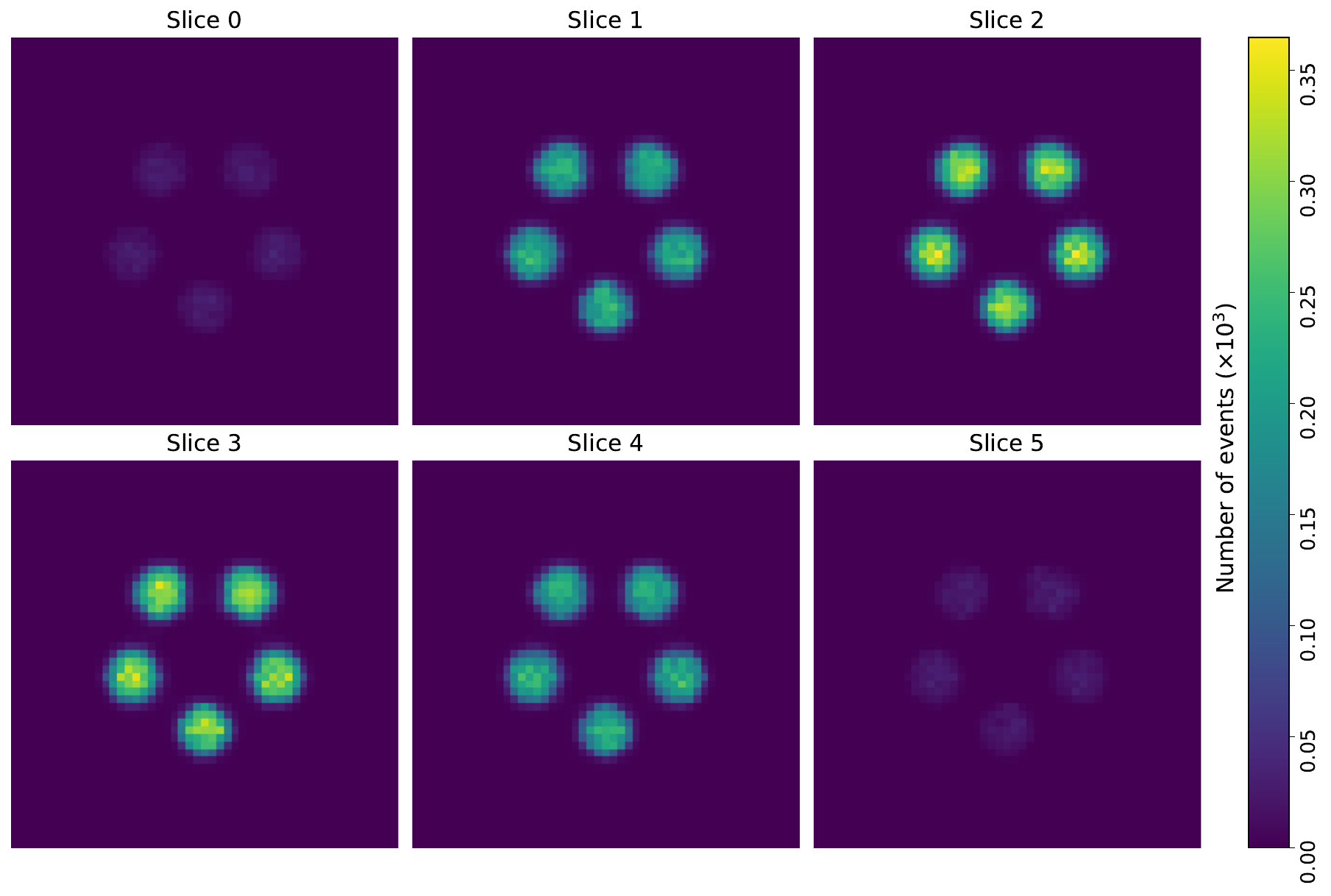}
\caption{
The recreated activity image closely matches the ground-truth distribution, with the reconstructed disks exhibiting the expected shape and intensity. Minor activity leakage is visible at the boundaries.
}
\label{fig:f_map_sim}
\end{figure}

\begin{figure}[h!]
\centering
\includegraphics[width=6.5in]{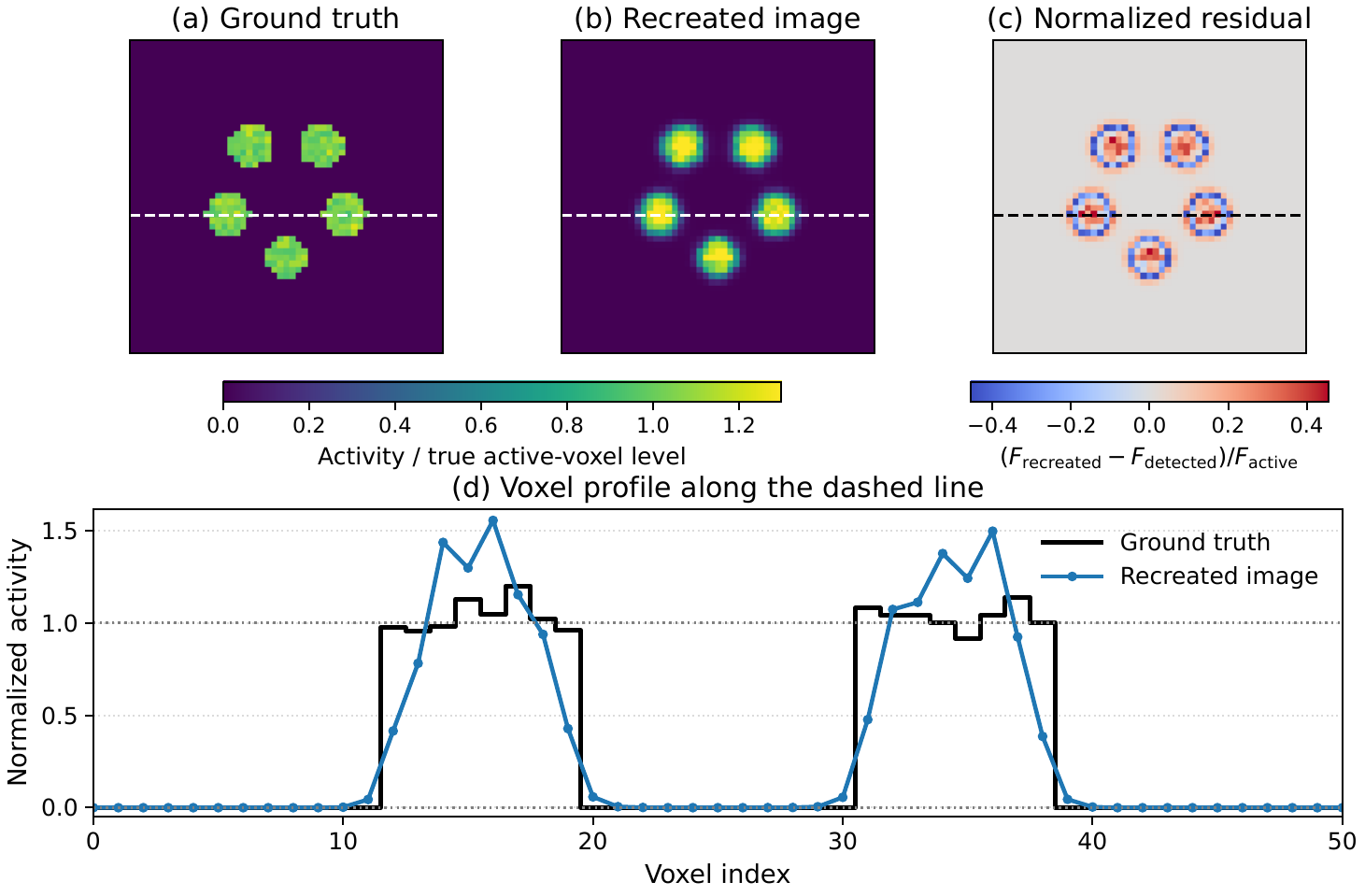}
\caption{
Comparison of the ground-truth and esimated activity distributions for the center slice. The normalized residual is defined relative to the true active-voxel level, and the line profile is taken along the dashed row.
}
\label{fig:f_map_sim_comp}
\end{figure}

The estimated detected-activity distribution reproduced the five active regions of the simulated phantom with the expected spatial arrangement and relative intensity, Figure \ref{fig:f_map_sim}. Most differences from the ground truth were concentrated near the boundaries of the active regions, as also seen in the normalized residuals and line profile in Figure \ref{fig:f_map_sim_comp}. Some axial spillover was observed into slices 0 and 5, which contain no activity in the ground truth. This is consistent with the boundary spillover observed elsewhere in the estimated activity distribution. Despite this leakage, the active regions remained well localized and provided suitable spatial responsibilities for the subsequent voxel-wise lifetime estimation.

\begin{figure}[h!]
\centering
\includegraphics[width=5in]{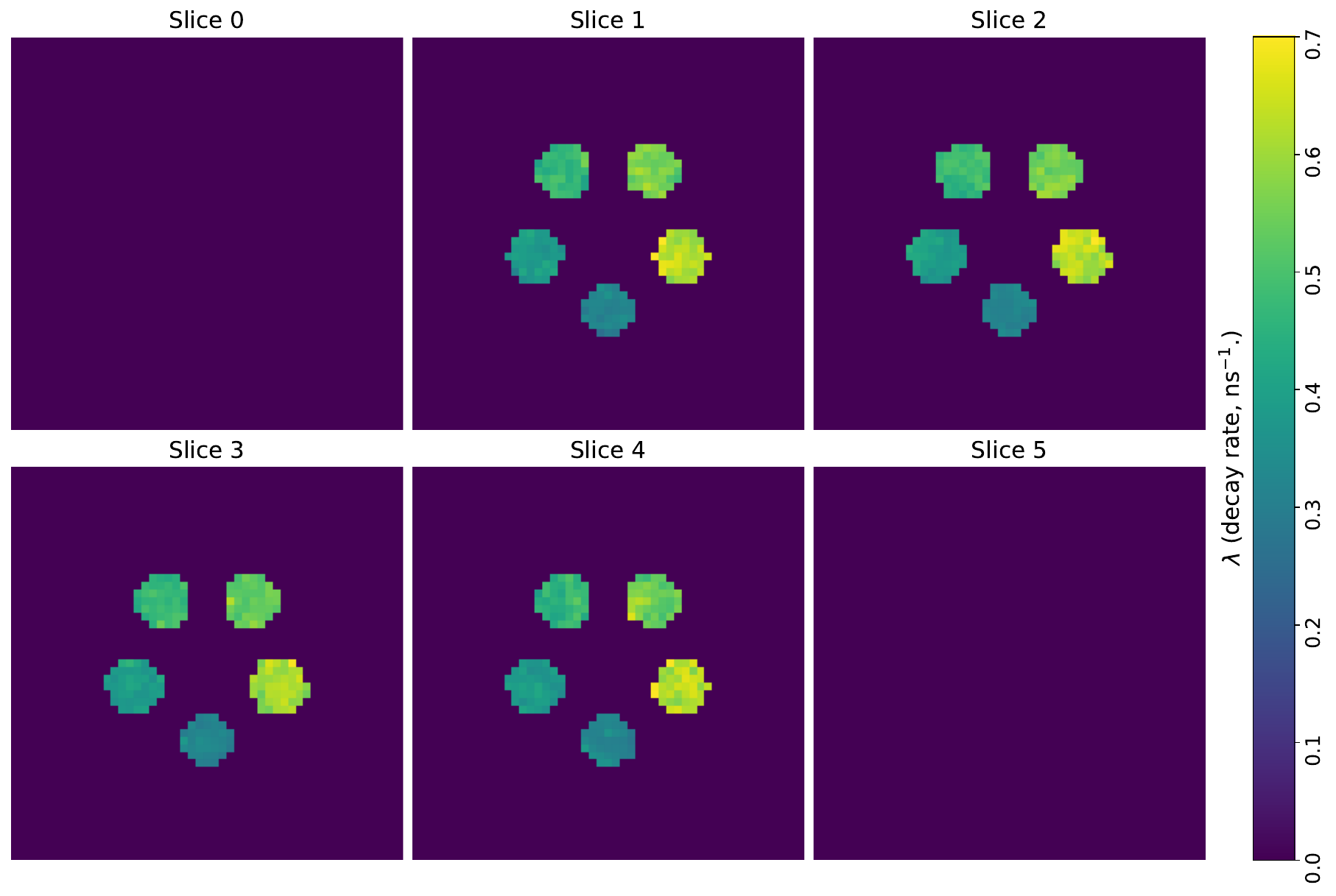}
\caption{
Reconstructed decay-rate map for the slow population. The five disks and their assigned decay rates are clearly recovered.
}
\label{fig:lambda_map_sim}
\end{figure}

\begin{figure}[H]
\centering
\includegraphics[width=5in]{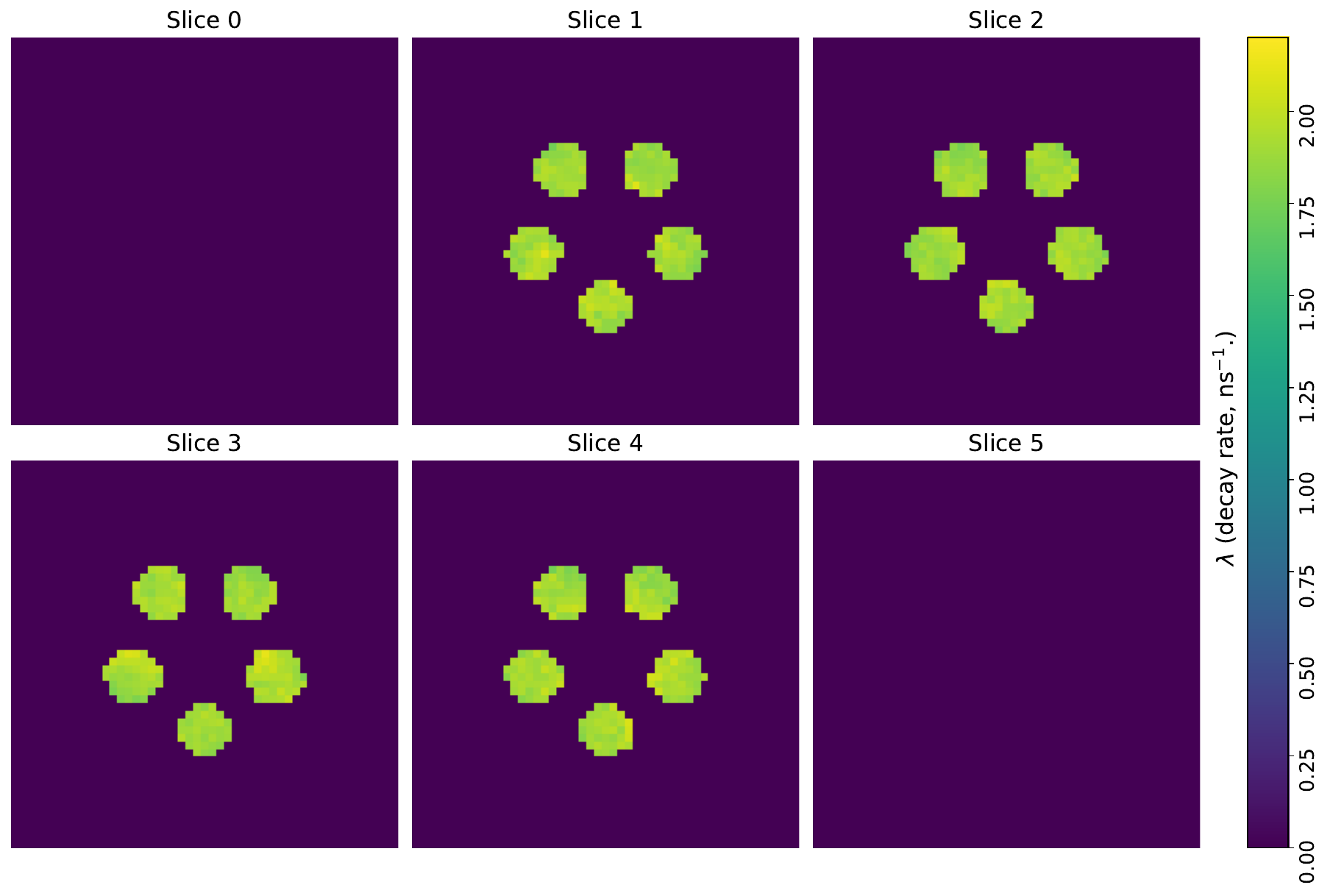}
\caption{
Reconstructed decay-rate map for the fast population. The five disks and their assigned decay rates are clearly recovered.
}
\label{fig:lambda_map_sim_fast}
\end{figure}

The estimated slow-population decay-rate map is shown in Figure \ref{fig:lambda_map_sim}. The five active regions remain spatially distinct, and the different decay-rate levels are recovered across the corresponding regions. Importantly, these values were not estimated from predefined regions of interest. A separate decay-rate posterior was obtained for each supported voxel using its population-specific event responsibilities. The resulting map therefore represents a voxel-wise estimate of the slow lifetime population rather than a regional fit applied to each disk. In contrast, the fast-population decay rate was assigned the same value throughout all five regions in the simulation. Accordingly, the estimated fast-population map in Figure \ref{fig:lambda_map_sim_fast} is nearly uniform across the active regions, with only small voxel-to-voxel variation. Together, these results show that the voxel-wise formulation can recover both spatially varying and spatially uniform population-specific decay rates.

\begin{figure}[H]
\centering
\includegraphics[width=6in]{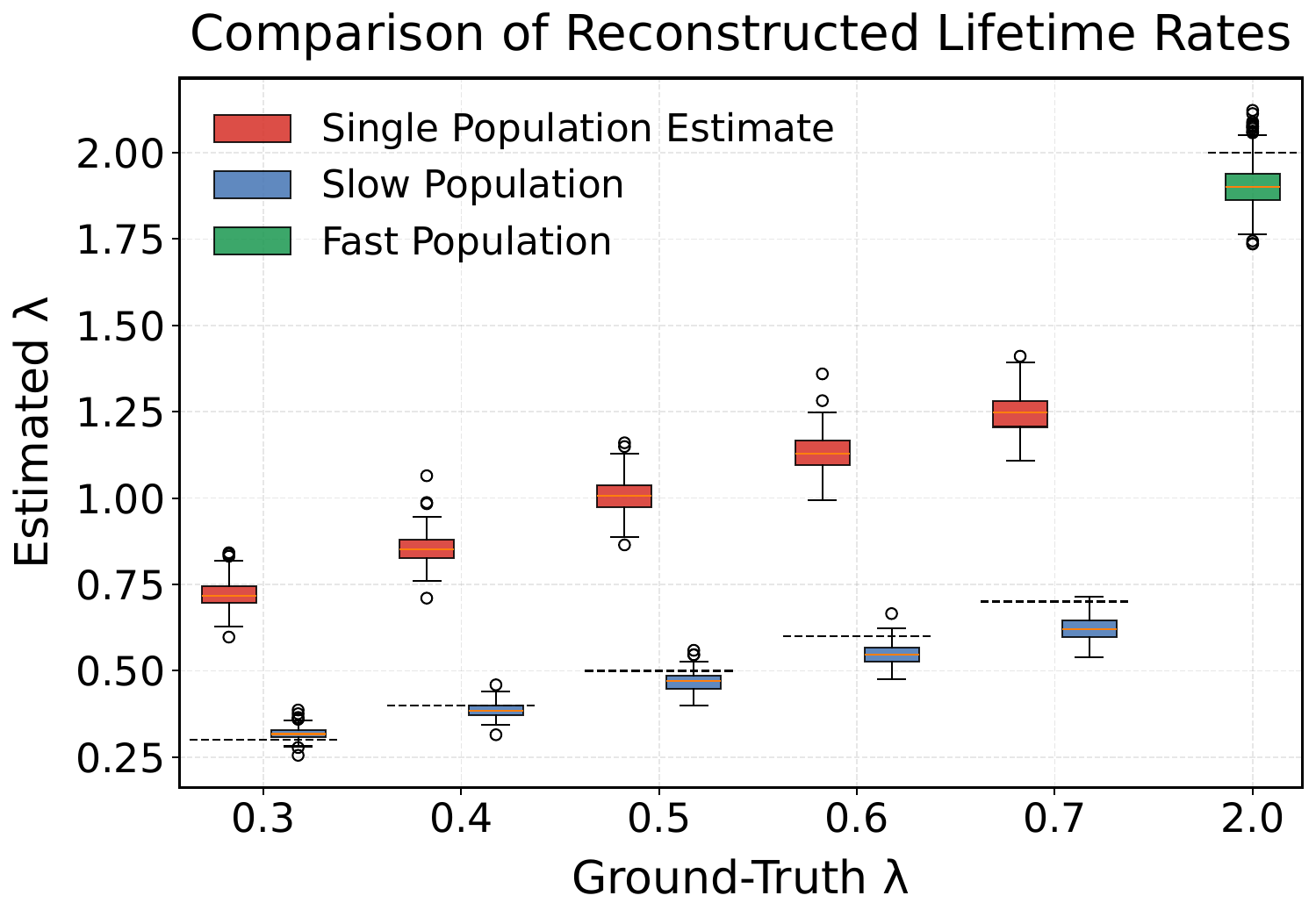}
\caption{
Comparison of the single-population and population-specific rate estimates. The single-population estimates are biased, while the two-population model accurately recovers the fast population and distinguishes the different slow-population rates. Black dashed lines indicate the ground-truth rate for each group.
}
\label{fig:lambda_map_comp}
\end{figure}

The quantitative distributions of the voxel-wise decay-rate estimates are shown in Figure \ref{fig:lambda_map_comp}. The single-population estimates are systematically higher than the ground-truth slow-population rates across all five regions. This behavior is expected because a single exponential must represent events generated from both the slow and fast populations, causing the estimated rate to be pulled toward the faster component. In contrast, separating the two populations substantially reduces this bias. The slow-population estimates remain close to their respective ground-truth values and preserve the ordering of all five decay rates, while the fast-population estimates form a narrow distribution close to the common ground-truth rate of $2.0~\mathrm{ns}^{-1}$.

A small bias remains in the population-specific estimates. The slow-population rate is slightly overestimated at the lowest ground-truth rate and becomes progressively underestimated for the higher rates, while the fast population is also estimated slightly below its ground-truth value. The increasing bias at higher slow-population rates is consistent with the greater overlap between the slow and fast lifetime distributions as their decay rates become closer. Because population responsibilities are assigned probabilistically, events in this overlapping region cannot be uniquely associated with either population. In addition, the population responsibilities are calculated using global population rates, whereas the simulated slow rate varies spatially. For the faster slow-population regions, shorter slow events are therefore more likely to receive fast-population responsibility, leaving a relatively longer-lived subset for the voxel-wise slow estimate and shifting the estimated decay rate downward. Despite this residual bias, the population-specific estimates are substantially closer to the ground truth than the single-population estimates and maintain clear separation between the five slow-population rates and the common fast-population rate.

\begin{figure}[H]
\centering
\includegraphics[width=6in]{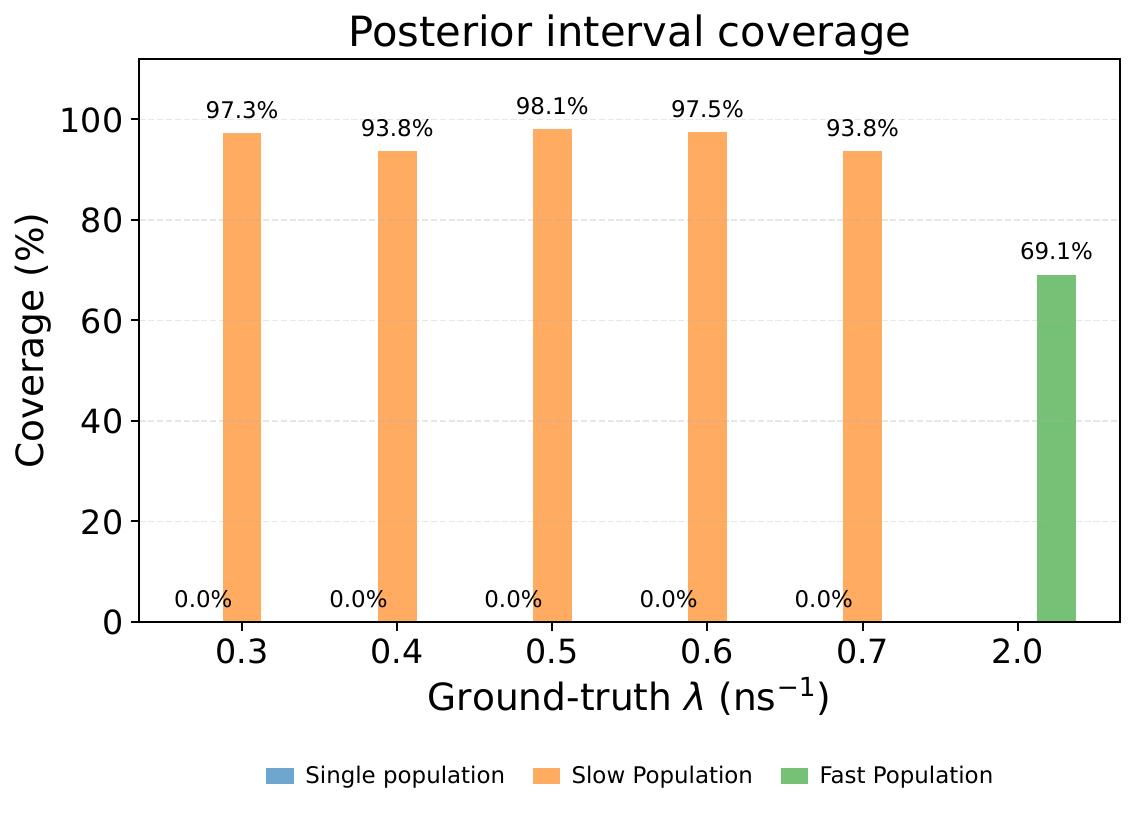}
\caption{Voxel-wise coverage of the estimated decay rates for the single-population and population-specific reconstructions. Coverage was defined as the percentage of voxels for which the ground-truth decay rate fell within two standard deviations of the estimated rate.}
\label{fig:coverage}
\end{figure}

The uncertainty estimates were further evaluated using empirical voxel coverage, defined as the percentage of voxels for which the ground-truth decay rate fell within two standard deviations of the estimated rate, Figure \ref{fig:coverage}. Two qualifications apply. First, the two-standard-deviation interval uses a Gaussian approximation to the Gamma posterior and may differ from an equal-tailed credible interval when the effective voxel count is small. Second, the percentage is calculated across voxels from a single noise realization, so it combines spatial and sampling variation and should not be interpreted as frequentist coverage. For the slow population, coverage remained consistently high across all five ground-truth rates, ranging from 93.8\% to 98.1\%. This indicates that, despite the small systematic biases observed in the voxel-wise estimates, the estimated uncertainty generally captured the true slow-population decay rate. In contrast, the single-population reconstruction achieved 0\% coverage for the slow-population ground truth across all five regions. This is consistent with the systematic bias observed in Figure \ref{fig:lambda_map_comp}, where the single-population estimates were shifted substantially toward the fast population.

Coverage for the fast population was lower, at 69.1\%. The reduced coverage is consistent with the small downward bias of the fast-population estimates observed in Figure \ref{fig:lambda_map_comp}. Because the voxel-wise distributions are relatively narrow, even a modest systematic shift can place the ground-truth value outside the estimated interval. These results show that the population-specific formulation provides substantially improved voxel-wise uncertainty characterization compared with the single-population model, while also indicating that the uncertainty of the fast-population estimates remains underestimated with some bias.

Finally, the conjugate lifetime estimation remained computationally inexpensive despite providing population-specific estimates at the voxel level. For the approximately $160\times10^{3}$ triple-coincidence events and $26\times26\times6$ reconstruction volume used in the simulation, estimation of the voxel-wise decay-rate posteriors required less than 10~s per population on a workstation equipped with an Intel Xeon Gold 6226R processor. No parallelization was used for this calculation. Since the conjugate update is performed independently for each voxel once the spatial and population responsibilities are available, the calculation is naturally parallelizable if additional computational performance is required. However, the short runtime obtained without parallelization indicates that voxel-wise posterior estimation itself introduces only a modest computational burden.

\subsection{Experimental Results}\label{sec:experimental_results}

\begin{figure}[H]
\centering
\includegraphics[width=5in]{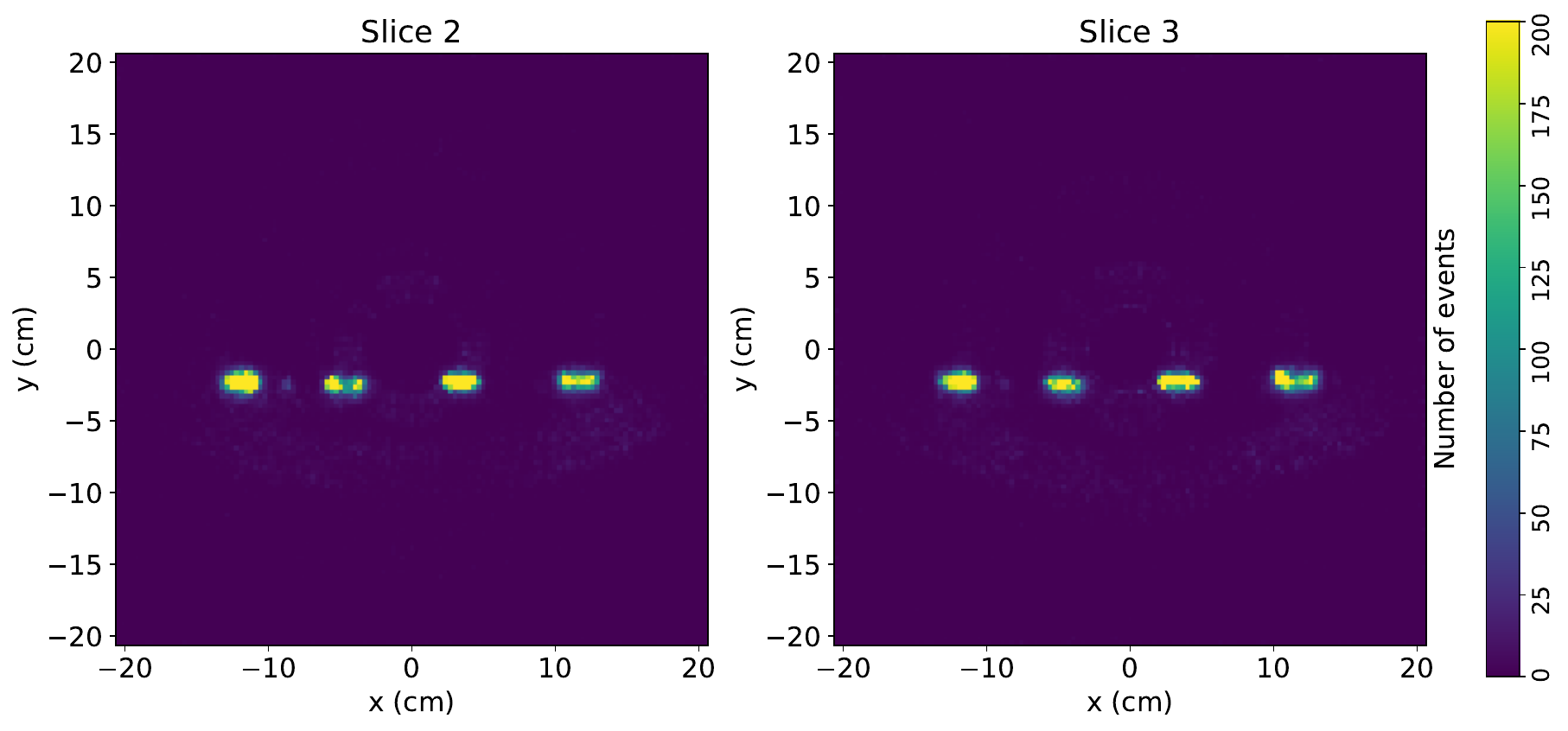}
\caption{
Central two slices of the reconstructed activity map, showing the four material samples.
}
\label{fig:f_map_exp}
\end{figure}

The estimated detected-activity distribution for the experimental dataset is shown in Figure \ref{fig:f_map_exp}. The four material samples are clearly localized within the reconstruction volume, with the estimated activity concentrated around the corresponding sample positions. As in the simulation, the activity image is used primarily to establish the spatial event-to-voxel responsibilities required for the subsequent lifetime estimation. The clear separation of the four active regions therefore provides the spatial basis for evaluating the population-specific voxel-wise lifetime estimates in the experimental data.

\begin{figure}[H]
\centering
\includegraphics[width=5in]{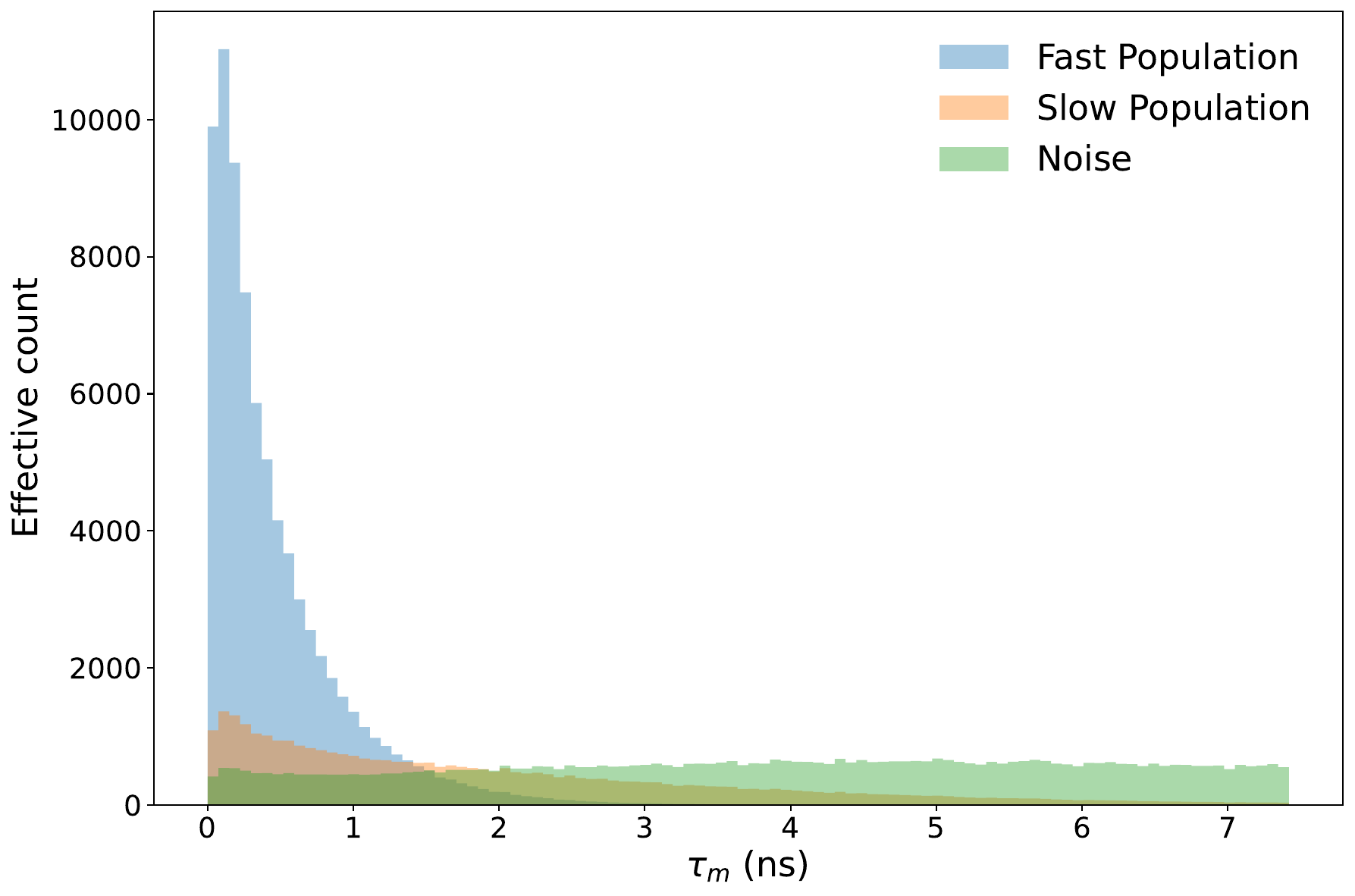}
\caption{
Effective slow, fast, and noise population counts as functions of the measured lifetime after soft population assignment.
}
\label{fig:pops_exp}
\end{figure}

The distribution of effective event counts after soft population assignment is shown in Figure \ref{fig:pops_exp}. The fast population dominates at short measured lifetimes and decreases rapidly as lifetime increases, while the slow population extends over a broader range of measured lifetimes. The noise contribution becomes relatively more important in the long-lifetime tail. Importantly, the population assignment is probabilistic rather than discrete. Each event is assigned a probability of originating from the slow, fast, or noise population based on its measured lifetime, and these probabilities are carried forward as fractional contributions in the voxel-wise estimation. Events in regions where the component distributions overlap therefore contribute to multiple populations according to their relative probabilities, rather than being forced into a single class by a fixed lifetime threshold.

The population-specific voxel-wise decay-rate estimates are shown in Figure \ref{fig:lambdas_exp}. Separate slow- and fast-population rates were estimated for each supported voxel, allowing differences among the four materials to be observed directly in the reconstructed maps. In the slow-population map, the first three metal samples show relatively similar decay-rate levels, whereas the quartz sample is clearly distinct and exhibits a substantially lower decay rate, corresponding to a longer-lived component. In the fast-population map, the three metal samples show similar decay-rate levels, while quartz again remains clearly separated from the metals. This behavior is consistent with the fast population carrying more material-specific information for the metals, whereas the slow population provides the strongest contrast for quartz. The spatial consistency of the estimates within each sample further shows that these differences are not limited to regional averages but are recovered across the individual voxels comprising each active region. The physical interpretation of these population-specific differences is discussed in more detail below.

\setcounter{subfigure}{0}
\begin{subfigure}[H]
\setcounter{subfigure}{0}
    \centering
    \begin{minipage}[b]{0.75\textwidth}
        \includegraphics[width=\linewidth]{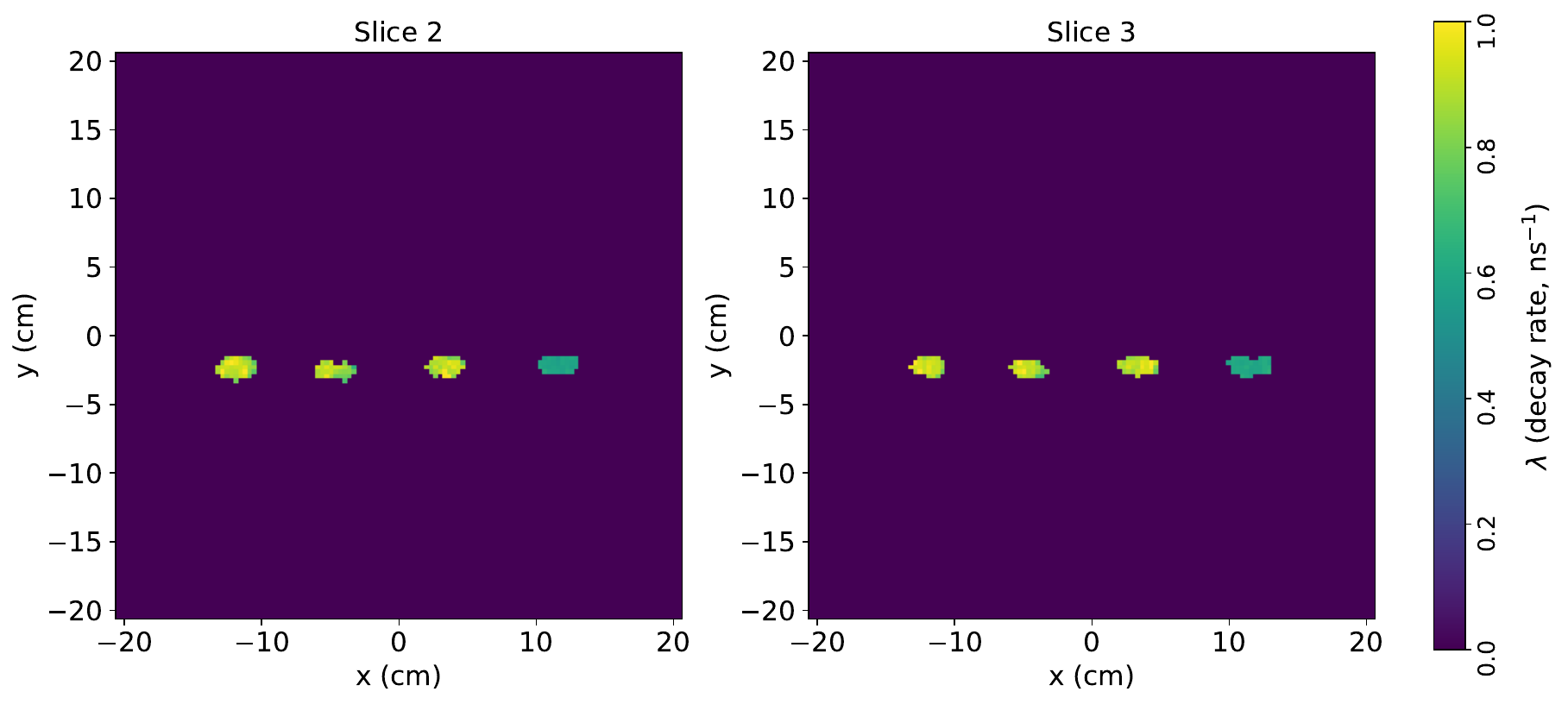}
        \caption{Estimated decay-rate map for the slow population.}
        \label{fig:Subfigure 1}
    \end{minipage}

\setcounter{subfigure}{1}
    \begin{minipage}[b]{0.75\textwidth}
        \includegraphics[width=\linewidth]{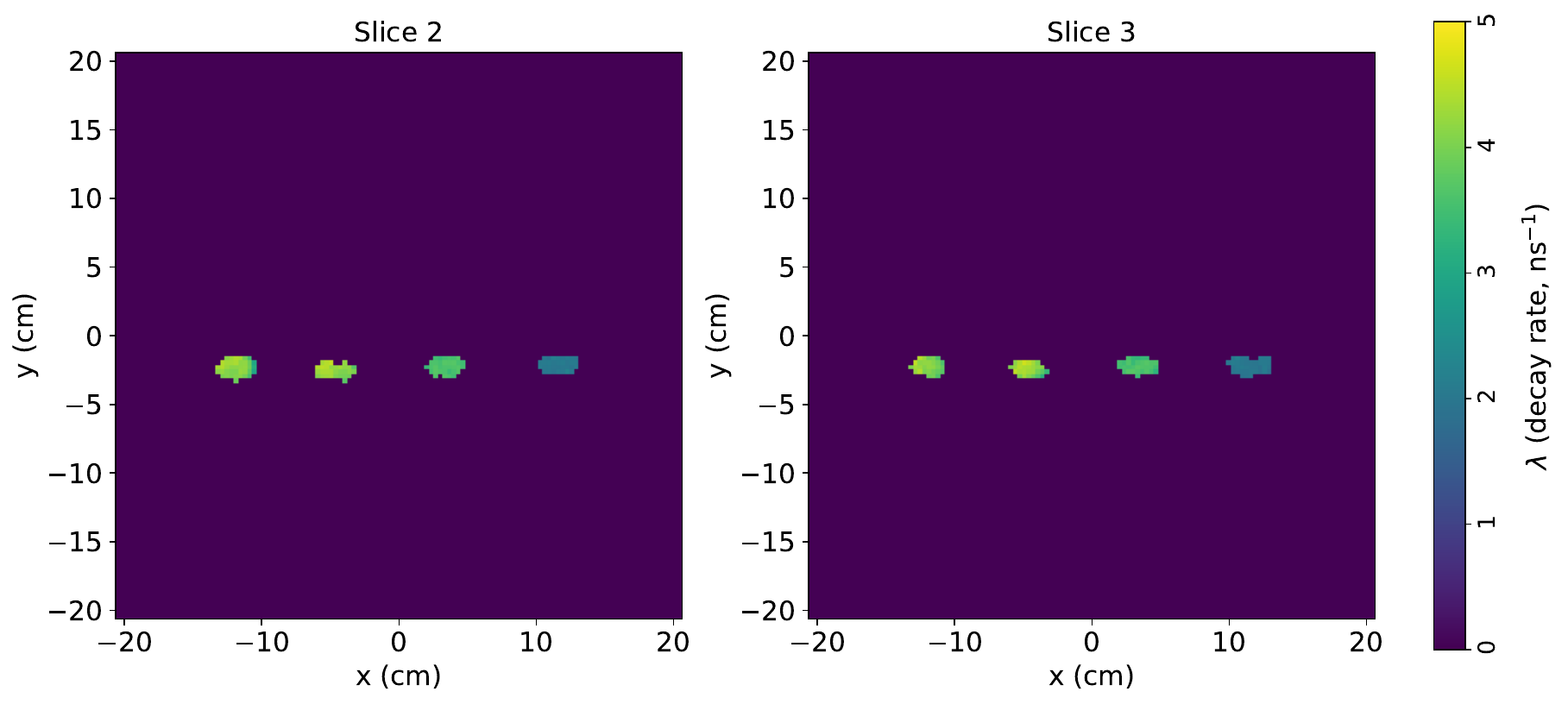}
        \caption{Estimated decay-rate map for the fast population.}
        \label{fig:Subfigure 2}
    \end{minipage}

\setcounter{subfigure}{-1}
    \caption{Estimated decay-rate maps for the two central slices of the experimental $^{124}$I dataset. Events were separated using soft population assignment before the population-specific conjugate reconstruction. \textbf{(a)} Slow-population rate map. \textbf{(b)} Fast-population rate map. Note that the two panels use different colour-bar ranges (0--1 and 0--5~\(\mathrm{ns}^{-1}\)) and are therefore not directly comparable. }
    \label{fig:lambdas_exp}
\end{subfigure}

\begin{figure}[H]
\centering
\includegraphics[width=5in]{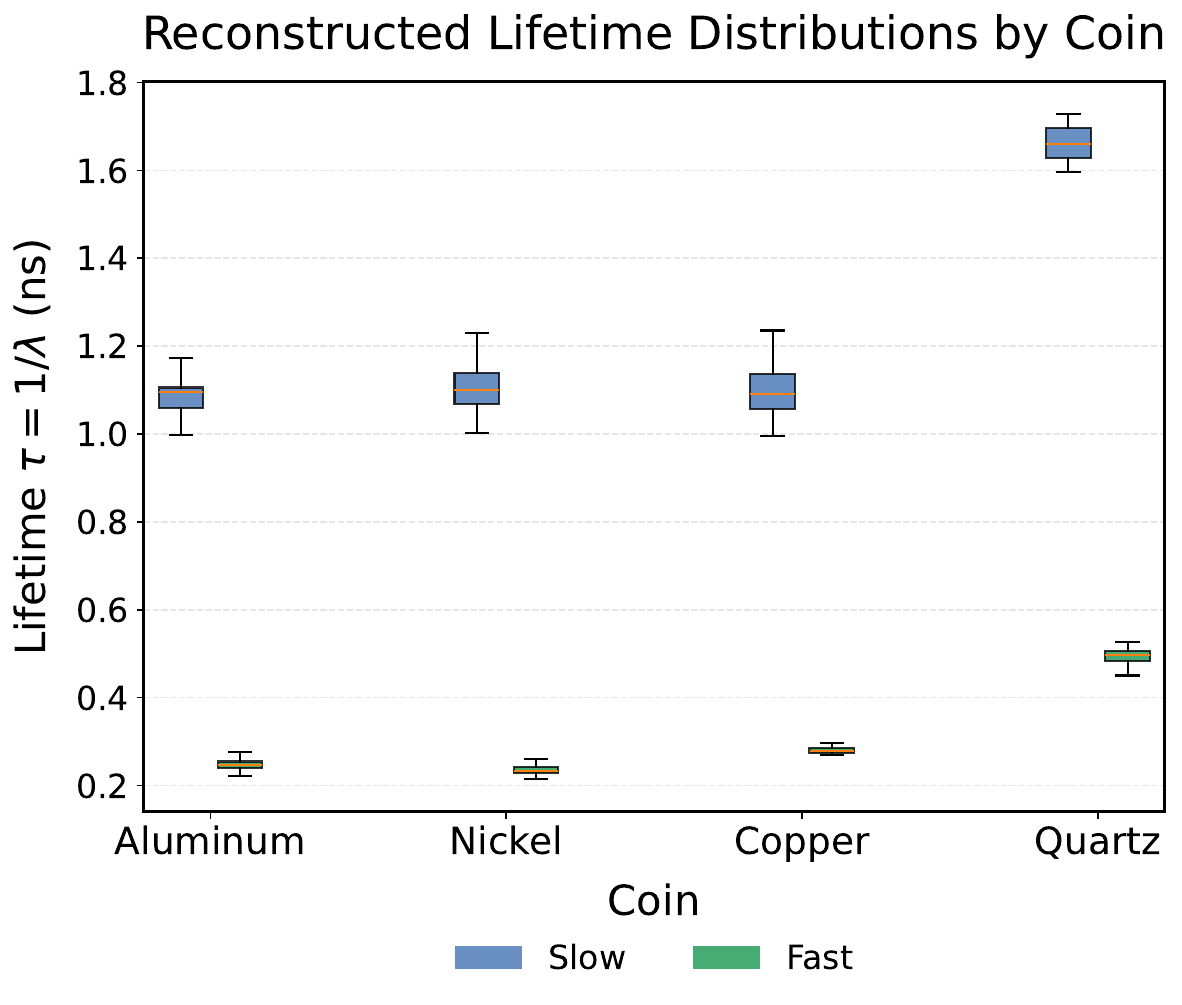}
\caption{
Reconstructed slow- and fast-population lifetime distributions for the aluminum, nickel, copper, and quartz samples. Quartz shows a distinctly longer slow-population lifetime.
}
\label{fig:lifetime_boxplot_exp}
\end{figure}

The material-wise distributions of the voxel-wise lifetime estimates are shown in Figure \ref{fig:lifetime_boxplot_exp}. Each distribution contains the lifetime estimates from the individual voxels belonging to the corresponding material region and therefore shows the spatial variation of the estimated lifetime within each sample. For the slow population, the three metal samples show strongly overlapping voxel distributions, with mean lifetimes of $\approx$1.09~ns for Al, 1.11~ns for Ni, and 1.10~ns for Cu. In contrast, the quartz distribution is clearly separated from the metals, with a mean lifetime of $\approx$1.66~ns. The fast population provides greater separation among the three metals, with mean lifetimes of $\approx$0.25~ns for Al, 0.24~ns for Ni, and 0.28~ns for Cu, while quartz again forms a distinct distribution with a mean lifetime of $\approx$0.50~ns. The relatively compact distributions within each material are also consistent with the spatially uniform regions observed in the voxel-wise lifetime maps. Given the 228~ps timing resolution and the nonnegative-lifetime truncation discussed below, the apparent inter-metal differences in the fast population should be regarded as tentative rather than fully resolved material distinctions.

The fast-population lifetimes in the three metal samples are consistent with a short-lived component dominated by direct positron annihilation. The estimated mean lifetimes were 0.25~ns for Al, 0.24~ns for Ni, and 0.28~ns for Cu. These values are somewhat longer than previously reported positron lifetimes in these materials, although the difference is relatively small compared with the temporal resolution of the measurement. Reported values include approximately 0.16~ns for Ni, 0.16--0.18~ns for Cu, and up to approximately 0.24~ns for vacancy-associated annihilation in Al. \cite{lapkiewicz_determining_2024, steinberger_positronium_2024} The experimental data used here have a timing resolution of 228~ps FWHM, which is comparable to the lifetimes of these fast processes. In addition, the present formulation uses only nonnegative measured lifetimes. For very short-lived events, timing uncertainty can shift part of the measured distribution below zero; removing these events truncates the left side of the distribution and can introduce an upward bias in the estimated fast-population lifetime. These effects may therefore contribute to the longer fast-population estimates observed in the metal samples.

The fast population in quartz should be interpreted more broadly. Its mean lifetime of 0.4951~ns cannot be assigned uniquely to a single annihilation process. At the available timing resolution, direct annihilation and the much shorter para-positronium contribution cannot be resolved reliably as separate voxel-wise populations. The estimated fast component in quartz should therefore be understood as an effective short-lived population containing contributions from both processes rather than as a direct measurement of either one individually.

The slow population in quartz has a clear physical interpretation. The mean voxel-wise lifetime of 1.6607~ns is consistent with the long-lived ortho-positronium component expected in quartz. Steinberger et al. previously analyzed the $^{124}$I quartz measurement using a regional lifetime spectrum and obtained an o-Ps lifetime of 1.624~ns, with a 68\% highest density intervals of 1.618--1.632~ns. \cite{steinberger_positronium_2024} The approximately 2.3\% difference between this value and the mean obtained here is small, particularly considering that the present result is obtained from independently estimated voxel-wise lifetimes rather than a single lifetime fit to the complete sample region. The close agreement therefore supports the interpretation of the reconstructed slow population in quartz as predominantly o-Ps. 

The slow population recovered in the metal samples requires a different interpretation, because positronium does not form in bulk metal. In a metal the conduction-electron gas screens the electron--positron Coulomb attraction over a Thomas--Fermi length comparable to or shorter than the 0.106~nm positronium Bohr radius, and Pauli blocking in the degenerate Fermi sea further suppresses formation of the bound state. Positrons in aluminum, nickel, and copper therefore annihilate directly with conduction electrons at a rate $\lambda = \pi r_e^2 c\, n_{\mathrm{eff}}$ set by the local effective electron density, where $\lambda$ is the annihilation rate (the reciprocal of the lifetime, $\tau = 1/\lambda$), $r_e \approx 2.82\times10^{-15}$~m is the classical electron radius (so that $\pi r_e^2$ sets the scale of the two-photon annihilation cross-section), $c$ is the speed of light, and $n_{\mathrm{eff}} = \gamma\, n_0$ is the effective electron density seen by the positron, equal to the average electron density $n_0$ enhanced by a factor $\gamma > 1$ that accounts for the positron attracting electrons toward itself. This yields intrinsic single-component lifetimes of approximately 0.11--0.16~ns and no long-lived tail. The fast-population estimates ($\approx$0.24--0.28~ns) are consistent with this direct-annihilation channel, broadened by the timing resolution. The slow component recovered in the metals ($\approx$1.1~ns) has no bulk-metal annihilation process. Since the reconstruction requires a slow signal fraction of 30\% in all materials (Equation~\ref{eq:population_mixture_proportions}), a slow class exists even in objects where there exist no such slow components, and in metals it is the resolution-broadened tail of the fast component as well as the real ortho-positronium events in the aqueous $^{124}$I solution in between the two disks where the bulk-metal annihilation is generated from the voxels of the metals. The $\approx$1.1~ns value is therefore an effective, model-induced long-lived component of the metals and is not a measured metal lifetime as in the case of the three metals. In this sense, the metal samples are a partial negative control: the framework returns a physically meaningful o-Ps lifetime for quartz while returning only a non-physical effective slow component for the metals, confirming that the method does not fabricate a genuine o-Ps signal where the electron density forbids one.

The slow population observed in the metal samples requires a different interpretation. The mean lifetimes of approximately 1.1~ns are substantially longer than reported bulk and defect-associated positron lifetimes in Al, Ni, and Cu, and therefore are unlikely to represent a distinct annihilation state within the bulk metals. Instead, this population should be regarded as an effective long-lived component of the measured lifetime distribution. Possible contributions include positronium formation at surfaces or interfaces, annihilations associated with the source environment surrounding the metal disks, and other long-lived events that are not individually resolved by the two-population model. In particular, the radioactive solution was placed between the paired metal disks, so events spatially associated with a metal sample need not originate exclusively from annihilation within the bulk metal. The present results therefore support a physical interpretation of the quartz slow population as predominantly o-Ps, while the slow component in the metals is retained as a statistical long-lived population without assigning it to a single annihilation mechanism.

Taken together, the experimental results demonstrate that the proposed framework can recover distinct lifetime behavior spatially at the voxel level from measured three-dimensional PLI data. The material-wise distributions reported above are formed from independently estimated voxel values rather than from a single lifetime fit over a predefined region. Each supported voxel therefore retains its own population-specific decay-rate estimate together with the corresponding Bayesian posterior information. This is an important distinction for future PLI applications, where spatial heterogeneity may itself carry diagnostic information and can be obscured when lifetime is summarized only at the regional level. The physical meaning of the estimated populations depends on the annihilation processes present in the material. In the present experiment, the long-lived quartz component is consistent with ortho-positronium, while the short-lived metal components are consistent with predominantly direct positron annihilation. Short-lived processes such as direct annihilation and para-positronium remain difficult to separate at the available timing resolution and may therefore contribute jointly to the estimated fast population. The purpose of the population-specific formulation is to recover these distinct lifetime scales spatially without requiring a separate regional fit for each material or annihilation component.

The voxel-wise lifetime estimation also remained computationally inexpensive for the experimental dataset. Using the same computational setup as in the simulation study, estimation of all supported voxel-wise decay-rate posteriors required less than 10~s per population without parallelization. Once the spatial and population responsibilities are available, the posterior calculation is independent across voxels and is therefore naturally parallelizable if additional computational performance is required for larger reconstruction volumes or datasets.

\section{Conclusion}

This work presents a population-specific conjugate Bayesian framework for three-dimensional positronium lifetime imaging with voxel-wise parameter estimation. The central contribution is the ability to estimate distinct lifetime populations independently at each supported voxel rather than assigning a single lifetime to a predefined region of interest. Each voxel retains its own population-specific decay-rate estimate together with Bayesian posterior information, providing both spatially resolved lifetime information and a direct measure of the uncertainty associated with the local estimate. The simulation study demonstrated that this voxel-wise formulation can recover spatially varying slow-population decay rates while simultaneously estimating a common fast population. In contrast, a single-population model produced systematically biased estimates because events generated by different lifetime populations were represented by a single exponential rate. Separating the populations substantially reduced this bias and preserved the differences among the simulated lifetime regions. The uncertainty analysis further showed high interval coverage for the slow-population estimates, demonstrating that the framework provides useful statistical information in addition to the reconstructed lifetime maps, while the fast-population coverage of 69.1\% indicates that the fast-population intervals remain too narrow at the present timing resolution. Application to experimental $^{124}$I data acquired on the Biograph Vision Quadra demonstrated the same framework on measured three-dimensional PLI data. Separate slow- and fast-population lifetime maps were obtained for aluminum, nickel, copper, and quartz without performing independent regional lifetime fits. The estimated populations showed material-dependent behavior, including a long-lived quartz component consistent with ortho-positronium and short-lived components in the metals consistent with predominantly direct positron annihilation.  More importantly, these results were obtained as distributions of individual voxel estimates, demonstrating that physically meaningful lifetime differences can be preserved spatially rather than reduced to a single value for each sample.

The conjugate formulation also makes voxel-wise Bayesian inference computationally practical. Once the spatial and population responsibilities are available, the posterior parameters are obtained directly from weighted sufficient statistics, avoiding numerical optimization and posterior sampling at individual voxels. For both the simulation and experimental datasets, estimation of all supported voxel-wise decay-rate posteriors required less than 10~s per population on the workstation used in this study, without parallelization. Since the voxel calculations are independent, the estimation is also naturally parallelizable for larger reconstruction volumes or datasets.

Together, these results establish a computationally efficient approach for population-specific, voxel-wise lifetime estimation with Bayesian uncertainty quantification in three-dimensional PLI. The ability to retain local lifetime information and its statistical uncertainty is particularly important for future diagnostic applications, where spatial heterogeneity may contain information that is lost through regional averaging. The proposed framework therefore provides a practical basis for extending PLI from regional lifetime measurements toward spatially resolved statistical imaging.

\section*{Conflict of Interest Statement}
The authors declare that the research was conducted in the absence of any commercial or financial relationships that could be construed as a potential conflict of interest.

\section*{Author Contributions}

\textbf{BU}: Conceptualization, Methodology, Software (voxel-wise Bayesian reconstruction framework), Visualization, Writing -- original draft. \textbf{NR}: Data curation, Software (independent list-mode processing and analysis pipeline, developed outside the vendor software), Investigation, Validation, Writing -- review \& editing. \textbf{GG:} Methodology, Software, Validation. \textbf{KaS} (Katrina Stephenson): Methodology, Investigation. \textbf{AR}: Resources, Supervision, Funding acquisition. \textbf{KuS} (Kuangyu Shi): Resources, Supervision, Validation. \textbf{HHH}: Conceptualization, Methodology, Project administration, Supervision, Writing -- review \& editing. All authors contributed to the article and approved the submitted version.

\section*{Funding}
H.-H.H. was supported in part by NSF grants DMS-1924792 and DMS-2318925. 

\section*{Acknowledgments}
The authors thank the collaboration behind Ref.~\cite{steinberger_positronium_2024} and the staff of the University Clinic for Nuclear Medicine, Inselspital Bern, for acquiring the $^{124}$I Biograph Vision Quadra dataset. We are also grateful to our colleagues at the School of Data, Mathematical, and Statistical Sciences, University of Central Florida; our collaborators at the Department of Biomedical Research (DBMR), University of Bern, and Inselspital; and the J-PET group for helpful internal discussions.

%Bibliography

\bibliographystyle{unsrt}  
\bibliography{references, reference}

\clearpage
\appendix

\setcounter{equation}{0}
\setcounter{figure}{0}
\setcounter{table}{0}

\renewcommand{\theequation}{A\arabic{equation}}
\renewcommand{\thefigure}{A\arabic{figure}}
\renewcommand{\thetable}{A\arabic{table}}

\pagenumbering{gobble}

\section{APPENDIX}

This document accompanies the main article and provides: (S1) a derivation of the assignment-uncertainty variance correction used in Section 2.4 of the main text from first principles, together with an explicit statement of when the first-order approximation is expected to be accurate; and (S2) a quantitative analysis of the interval-coverage results reported in Section 3.1, isolating how much of the observed fast-population coverage shortfall (69.1\% against a nominal $\sim$95.4\% two-standard-deviation target) can plausibly be attributed to interval shape, estimator bias, or variance underestimation, with a concrete recommended diagnostic.

\subsection{Derivation of the Assignment-Uncertainty Variance Correction}
\label{sec:s1}

Recall from the main text that, for population $z\in\{s,f\}$ and voxel $j$, the reconstructed decay rate is the posterior mean of a $\mathrm{Gamma}(\alpha_{0,z}+n_{j,z},\,\beta_{0,z}+S_{j,z})$ distribution,
\begin{equation}
\hat\lambda_{j,z} = g(n_{j,z}, S_{j,z}) \equiv \frac{\alpha_{0,z}+n_{j,z}}{\beta_{0,z}+S_{j,z}},
\label{eq:s_post_mean}
\end{equation}
where $n_{j,z}=\sum_k w_{k,j}^{(z)}$ and $S_{j,z}=\sum_k w_{k,j}^{(z)}\tau_k$ are computed from the soft (expected) assignment weights $w_{k,j}^{(z)} = \mathbb{E}[A_{k,j,z}\mid\mathcal{D}]$, and $A_{k,j,z}\in\{0,1\}$ is the (unobserved) true joint voxel-population indicator for event $k$.

\subsubsection{Law of total variance}

Because $w_{k,j}^{(z)}$ is an expectation, not a realization, the quantities $n_{j,z}$ and $S_{j,z}$ computed from the \emph{true} indicators $A_{k,j,z}$ would themselves be random even conditional on all other information; treating $w_{k,j}^{(z)}$ as fixed and plugging it directly into the Gamma posterior variance (Eq.~54 of the main text) accounts for only part of the total uncertainty in $\hat\lambda_{j,z}$. Writing $\mathbf{A}=\{A_{k,j,z}\}_k$ for the full latent assignment vector at voxel $j$, population $z$, the law of total variance gives the decomposition used implicitly in the main text,
\begin{equation}
\mathrm{Var}(\hat\lambda_{j,z}) \;=\; \underbrace{\mathbb{E}\bigl[\mathrm{Var}(\hat\lambda_{j,z}\mid\mathbf{A})\bigr]}_{\text{within-assignment variance}} \;+\; \underbrace{\mathrm{Var}\bigl(\mathbb{E}[\hat\lambda_{j,z}\mid\mathbf{A}]\bigr)}_{\text{between-assignment variance}}.
\label{eq:s_totvar}
\end{equation}
The main text approximates the first term by $\mathrm{Var}_{\mathrm{Gamma}}$ (Eq.~54) evaluated at the plug-in estimates $(n_{j,z}, S_{j,z})$ rather than averaged over the distribution of $\mathbf{A}$, and approximates the second term -- the variance of the posterior mean induced by not knowing $\mathbf{A}$ exactly -- by a first-order (delta-method) expansion of $g$, which we re-derive here explicitly.

\subsubsection{Delta-method expansion}

Since $n_{j,z}=\sum_k A_{k,j,z}$ and $S_{j,z}=\sum_k A_{k,j,z}\tau_k$ are themselves sums of independent Bernoulli$(w_{k,j}^{(z)})$-weighted terms under the working model $A_{k,j,z}\sim\mathrm{Bernoulli}(w_{k,j}^{(z)})$, a first-order Taylor expansion of $g(n,S)$ around $(n_{j,z}, S_{j,z})$ gives
\begin{equation}
\mathrm{Var}\bigl(g(n,S)\bigr) \approx \left(\frac{\partial g}{\partial n}\right)^{\!2}\mathrm{Var}(n) + \left(\frac{\partial g}{\partial S}\right)^{\!2}\mathrm{Var}(S) + 2\frac{\partial g}{\partial n}\frac{\partial g}{\partial S}\,\mathrm{Cov}(n,S),
\label{eq:s_delta}
\end{equation}
a standard multivariate delta-method result \citeapp{casella_statistical_2002}. Direct differentiation gives $\partial g/\partial n = 1/(\beta_{0,z}+S_{j,z})$ and $\partial g/\partial S = -(\alpha_{0,z}+n_{j,z})/(\beta_{0,z}+S_{j,z})^2$; substituting into Eq.~\eqref{eq:s_delta} together with $\mathrm{Var}(n_{j,z})$, $\mathrm{Var}(S_{j,z})$, and $\mathrm{Cov}(n_{j,z},S_{j,z})$ from the Bernoulli model (Eqs.~57--59 of the main text) reproduces Eq.~60 of the main text term for term. This confirms that the main text's correction is precisely the delta-method approximation of the between-assignment term of Eq.~\eqref{eq:s_totvar}, combined with a plug-in (rather than fully marginalized) approximation of the within-assignment term.

\subsubsection{Validity conditions and expected failure modes}

Two approximations are made in obtaining Eq.~61 of the main text, and both are only first-order accurate:

\begin{proposition}[Validity of the first-order correction]
\label{prop:s_validity}
The approximation $\mathrm{Var}(\hat\lambda_{j,z}) \approx \mathrm{Var}_{\mathrm{Gamma}} + \mathrm{Var}_{\mathrm{assign}}$ neglects two classes of terms: (i) the curvature of $\mathrm{Var}_{\mathrm{Gamma}}(n,S)$ itself over the range of $(n,S)$ induced by the uncertainty in $\mathbf{A}$ (a Jensen-type gap between $\mathbb{E}[\mathrm{Var}_{\mathrm{Gamma}}(n,S)]$ and $\mathrm{Var}_{\mathrm{Gamma}}(\mathbb{E}[n],\mathbb{E}[S])$), and (ii) third- and higher-order terms in the Taylor expansion of $g$ used to obtain Eq.~\eqref{eq:s_delta}. Both omitted terms are controlled by the relative dispersion of $(n_{j,z}, S_{j,z})$, which by the Bernoulli model is largest when many individual event weights $w_{k,j}^{(z)}$ are close to $0.5$ (maximal per-event assignment ambiguity, where $w(1-w)$ is maximized) rather than close to $0$ or $1$.
\end{proposition}

\begin{remark}
This gives a concrete, checkable prediction: populations or voxels where many events have near-ambiguous assignment probabilities (i.e., where the slow/fast/noise lifetime distributions overlap substantially in the relevant $\tau$ range) are exactly where the first-order correction is expected to be least reliable. The fast population, which by construction carries the majority (0.70) of the signal fraction and therefore overlaps with both the slow and noise components across a wide range of $\tau$ (see Fig.~10 of the main text, where the fast, slow, and noise densities visibly overlap through much of the $0$--$2$ ns range), is a plausible candidate for exactly this failure mode, independently of any coding error.
\end{remark}

This does not by itself establish that the first-order approximation is the sole or primary cause of the fast-population undercoverage reported in the main text (Section 3.1); Section~\ref{sec:s2} below shows that a moderate estimator bias could equally well explain the same numeric shortfall, and the two explanations are not mutually exclusive. We recommend the direct diagnostics of Remark~\ref{rem:s_diagnostic} below to distinguish them empirically.

\subsection{Quantitative Analysis of the Reported Interval Coverage}
\label{sec:s2}

\subsubsection{Is the coverage shortfall a Gaussian-approximation artifact?}

The main text reports coverage using a $\pm 2$ (posterior) standard-deviation interval around the posterior mean. Because the underlying posterior is Gamma-distributed (right-skewed, especially for small shape parameter), one might suspect that approximating it by a symmetric Gaussian interval is itself responsible for miscoverage. We checked this directly: for a $\mathrm{Gamma}(\alpha,\beta)$ distribution, Table~\ref{tab:s1} compares the naive interval $[\text{mean}-2\,\text{sd},\,\text{mean}+2\,\text{sd}]$ (clipped at zero, since a decay rate cannot be negative) against the exact $2.5\%$--$97.5\%$ equal-tailed quantile interval, and reports the \emph{actual} coverage probability that the naive interval achieves under the true Gamma distribution.

\begin{table}[h]
\centering
\caption{Naive Gaussian vs.\ exact Gamma intervals, and the true coverage of the naive interval, across a range of shape parameters $\alpha$ (rate parameter $\beta=1$; results are scale-invariant). ``Naive coverage'' is $P(\text{naive interval covers the true value})$ under $X\sim\mathrm{Gamma}(\alpha,\beta)$.}
\label{tab:s1}
\begin{tabular}{rrrrrrr}
\toprule
$\alpha$ & mean & sd & naive lo & naive hi & exact $2.5\%$ & naive coverage \\
\midrule
2   & 2.00  & 1.41 & 0.00 (clipped) & 4.83   & 0.24  & 95.3\% \\
4   & 4.00  & 2.00 & 0.00           & 8.00   & 1.09  & 95.8\% \\
8   & 8.00  & 2.83 & 2.34           & 13.66  & 3.45  & 95.9\% \\
20  & 20.00 & 4.47 & 11.06          & 28.94  & 12.22 & 95.7\% \\
100 & 100.00& 10.00& 80.00          & 120.00 & 81.36 & 95.5\% \\
\bottomrule
\end{tabular}
\end{table}

Table~\ref{tab:s1} shows that the naive Gaussian interval's true coverage remains close to its nominal $95.4\%$ target across the full range of shape parameters tested, including quite small $\alpha$. \textbf{This rules out Gaussian/Gamma shape mismatch as the primary explanation for the observed 69.1\% fast-population coverage}: a symmetric-interval approximation to a skewed posterior is, on its own, a second-order effect here, not a first-order one. (We nonetheless recommend reporting exact equal-tailed or highest-posterior-density Gamma intervals rather than mean-$\pm2$-SD in future work, both because it removes any risk of an invalid negative lower bound and because it is only marginally more work to compute in closed form for a Gamma posterior.)

\subsubsection{Quantifying the bias/variance explanation}
\label{sec:s2b}

If shape mismatch is not the driver, the two remaining candidate explanations are (i) a systematic bias in the point estimate itself, and (ii) an underestimated posterior standard deviation (i.e., the correction of Section~\ref{sec:s1} not being large enough). Table~\ref{tab:s2} quantifies each in isolation, using a local Gaussian approximation valid near the mean (reasonable for the fast population's larger effective counts): if the estimator is biased low by $\delta$, expressed in units of its own reported standard deviation $\sigma$, the coverage of the reported $[\hat\lambda-2\sigma,\hat\lambda+2\sigma]$ interval for the true value is $\Phi(2+\delta/\sigma)-\Phi(-2+\delta/\sigma)$; if instead the estimator is unbiased but the true standard deviation is larger than reported by a factor $f$, the coverage is $\Phi(2/f)-\Phi(-2/f)$, where $\Phi$ is the standard normal CDF.

\begin{table}[h]
\centering
\caption{Coverage of a reported 2-SD interval under (left) a pure bias of $\delta/\sigma$ standard deviations, holding the interval width fixed, or (right) a pure variance underestimate by a factor $f$, holding the estimator unbiased.}
\label{tab:s2}
\begin{tabular}{rr @{\hspace{2.5em}} rr}
\toprule
$\delta/\sigma$ & coverage & $f$ & coverage \\
\midrule
0.00 & 95.4\% & 1.0 & 95.4\% \\
0.50 & 92.7\% & 1.2 & 90.4\% \\
1.00 & 84.0\% & 1.4 & 84.7\% \\
\textbf{1.50} & \textbf{69.1\%} & 1.6 & 78.9\% \\
1.75 & 59.9\% & \textbf{2.0} & \textbf{68.3\%} \\
2.00 & 50.0\% & 2.5 & 57.6\% \\
\bottomrule
\end{tabular}
\end{table}

The observed 69.1\% fast-population coverage is reproduced almost exactly by either (a) a systematic bias of about $1.5\sigma$ with an otherwise correctly-sized interval, or (b) a correctly-centered interval whose reported standard deviation is too small by a factor of about 2, or by intermediate combinations of smaller bias and smaller variance underestimation along the curve connecting these two extremes.

\begin{remark}[Recommended diagnostic]
\label{rem:s_diagnostic}
These two explanations are directly distinguishable from data already computed in the pipeline, without new experiments: (1) compute, for the fast-population voxels, the empirical bias $\hat\lambda_{j,f}-\lambda_{f,\mathrm{true}}$ in units of the reported posterior SD, and compare its typical magnitude against the $1.5\sigma$ benchmark above; (2) recompute the reported coverage using $\mathrm{Var}_{\mathrm{Gamma}}$ alone (Eq.~54 of the main text, i.e.\ omitting the assignment-uncertainty correction) versus the full corrected variance (Eq.~61); if coverage barely changes between the two, the shortfall is predominantly a bias effect and the assignment-uncertainty correction is not the limiting factor; if coverage improves substantially with the correction but still falls well short of nominal, the correction is directionally right but insufficient in magnitude (consistent with Proposition~\ref{prop:s_validity}'s prediction that the first-order approximation degrades precisely where fast-population events have high assignment ambiguity). Either outcome is informative and low-cost to obtain, since both quantities are already computed internally by the reconstruction.
\end{remark}

\subsubsection{Toward a more comprehensive validation: simulation-based calibration}

The coverage metric reported in the main text -- the fraction of voxels whose ground truth falls within a single nominal interval -- is informative but is a single summary statistic evaluated at one nominal level (95.4\%, via $\pm2$ SD) and is specific to the particular simulated ground-truth values used. A more comprehensive check of posterior calibration, standard in the modern Bayesian workflow literature, is simulation-based calibration (SBC) \citeapp{talts_validating_2018}: repeatedly (a) drawing a parameter value from the prior, (b) simulating data from the forward model at that parameter value, (c) fitting the posterior, and (d) computing the rank of the true parameter within posterior draws (or, for a fully analytic posterior as here, within the CDF of the fitted Gamma posterior). Under a correctly calibrated posterior, these ranks (or CDF values) should be uniformly distributed across many repetitions; systematic deviations from uniformity (e.g., ranks concentrated away from the tails, indicating overconfidence) would directly diagnose the miscalibration seen here across a full range of nominal coverage levels rather than only at the $\pm2$-SD level, and would do so per-population, allowing direct comparison of the slow- and fast-population calibration curves. We recommend this as a natural extension for validating the corrected posterior variance of Eq.~61 in future work.

\bibliographystyleapp{unsrt}
\bibliographyapp{references,reference}

\end{document}